\documentclass{egpubl}
\usepackage{eurovis2021}     

\usepackage[T1]{fontenc}
\usepackage{dfadobe}  

\usepackage{cite}  
\BibtexOrBiblatex
\electronicVersion
\PrintedOrElectronic
\ifpdf \usepackage[pdftex]{graphicx} \pdfcompresslevel=9
\else \usepackage[dvips]{graphicx} \fi

\usepackage[T1]{fontenc}
\usepackage{dfadobe}
\usepackage{subfigure}
\usepackage{amsmath}
\usepackage{amssymb}
\usepackage{appendix}
\usepackage{booktabs}
\usepackage[svgnames,table]{xcolor}
\usepackage{xspace}
\usepackage{dblfloatfix}
\usepackage[version=4]{mhchem}
\usepackage{multirow}

\usepackage{egweblnk}
\newcommand{\etal}{\textit{et al.}\xspace}

\title[Visual Analysis of Electronic Densities and Transitions in Molecules]
      {Visual Analysis of Electronic Densities and Transitions in Molecules}
\author[Talha Bin Masood et al.]
    { 
    \parbox{0.75\textwidth} 
    {\centering 
        Talha Bin Masood
        \thanks{\{talha.bin.masood | signe.sidwall.thygesen | mathieu.linares | alexei.abrikossov | ingrid.hotz\}@liu.se}
        $^{1}$\orcid{0000-0001-5352-1086},
        Signe Sidwall Thygesen
        $^{1}$\orcid{0000-0001-9156-647X},
        Mathieu Linares
        $^{1,2}$\orcid{0000-0002-9720-5429},
        Alexei I. Abrikosov
        $^{1}$\orcid{0000-0002-3322-0646},
        Vijay Natarajan
        \thanks{vijayn@iisc.ac.in}
        $^{3}$\orcid{0000-0002-7956-1470}, and
        Ingrid Hotz
       $^{1}$\orcid{0000-0001-7285-0483}
    }
    \\
    {\parbox{\textwidth}{\centering 
        $^1$Department of Science and Technology (ITN), Link\"oping University, Norrk\"oping, Sweden\\
        $^2$Laboratory of Organic Electronics, Link\"oping University, Norrk\"oping, Sweden\\
        $^3$Indian Institute of Science, Bangalore, India}
    }
}

\begin{document}

%\teaser{
% \includegraphics[width=\linewidth]{eg_new}
% \centering
%  \caption{New EG Logo}
%\label{fig:teaser}
%}

\maketitle

\begin{abstract}
The study of electronic transitions within a molecule connected to the absorption or emission of light is a common task in the process of the design of new materials. The transitions are complex quantum mechanical processes and a detailed analysis requires a breakdown of these processes into components that can be interpreted via characteristic chemical properties.
We approach these tasks by providing a detailed analysis of the electron density field. This entails methods to quantify and visualize electron localization and transfer from molecular subgroups combining spatial and abstract representations. The core of our method uses geometric segmentation of the electronic density field coupled with a graph-theoretic formulation of charge transfer between molecular subgroups. The design of the methods has been guided by the goal of providing a generic and objective analysis following fundamental concepts. We illustrate the proposed approach using several case studies involving the study of electronic transitions in different molecular systems. 
%-------------------------------------------------------------------------
%  ACM CCS 1998
%  (see https://www.acm.org/publications/computing-classification-system/1998)
% \begin{classification} % according to https://www.acm.org/publications/computing-classification-system/1998
% \CCScat{Computer Graphics}{I.3.3}{Picture/Image Generation}{Line and curve generation}
% \end{classification}
%-------------------------------------------------------------------------
%  ACM CCS 2012
%   (see https://www.acm.org/publications/class-2012)
%The tool at \url{http://dl.acm.org/ccs.cfm} can be used to generate
% CCS codes.



\end{abstract}
%-------------------------------------------------------------------------
\section{Introduction}
Molecular spectroscopy, dealing with the absorption or emission of light, plays an important role in material and biochemical applications. Its study is a common task when analyzing the chemical and physical properties of organic materials with a wide variety of technical applications, e.g., the design of new materials for organic solar cells.
Absorption and emission of light are related to electronic transitions, which involve the promotion of electrons from one state to another by absorbing or emitting photons~\cite{Kim2019}. 
The complex quantum mechanical process behind these transitions can be numerically calculated using modern quantum chemistry methods such as Density Functional Theory~(DFT).

Visualization has traditionally played an important role in analyzing the resulting electron density fields~\cite{Stone2011}. Statistical plots and energy diagrams are typically used, but spatial representations of the density field of a specific electronic state of the molecule are also studied. 
The spatial electron density distribution is often visualized using isosurfaces. An isovalue that `best’ depicts the system is manually chosen. This approach requires detailed knowledge about the molecule and localization of the electrons~\cite{Haranczyk2008}. As one can imagine, this approach can be cumbersome, time-consuming, and lacks any quantifiable information, thereby making comparisons across different molecules often difficult and sometimes impossible. 
A tool to quickly identify the character of each excitation and to facilitate the analysis of a series of molecules is missing. 

In this work, we propose an approach that automates the quantification and visualization of key measures for electron localization and transfer from molecular subgroups.
From a data point of view, this is a scalar field analysis and visualization problem. The design of our visual analysis framework is guided by the goal to develop a generic and efficient pipeline (Fig.~\ref{fig:overview}) that builds on fundamental concepts to generate effective and easy to interpret visualizations. More specifically, this includes (i)~designing a simple yet powerful partitioning algorithm that is chemically plausible, efficiently computable, and easy to communicate, (ii)~quantifying the field transition by solving a constrained optimization problem respecting simple rules, and (iii)~providing a visualization that is easy to understand and capable of encoding the characteristics of the nature of the electronic transition.

Our contributions can be summarized as:
\begin{itemize}
\item Formulation of the charge transition problem as a general partitioning and constraint optimization problem.
\item Introduction of a new automated method for quantitative analysis and comparison of charge distributions and transitions in molecular excitations.
\item Design and development of a framework for the visual analysis of electronic transitions in a series of molecules. 
\item Demonstration of the utility and significance of the framework via four case studies on molecules and metal complexes. 
\end{itemize}

\begin{figure*}[!tb]
    \centering
    \begin{tabular}{c@{}c@{}c@{}c@{}c@{}c}
    {\tiny CPK colors} 
    & {\tiny THIO} \includegraphics[width=0.1\columnwidth]{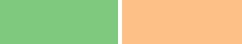} {\tiny QUIN} 
    &{\tiny Min} \includegraphics[width=0.1\columnwidth]{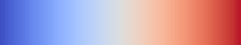} {\tiny Max} 
    & {\tiny $-v$} \includegraphics[width=0.1\columnwidth]{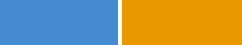} {\tiny $v$} 
    &{\tiny Min} \includegraphics[width=0.1\columnwidth]{Figures/blue_red_diverging.png} {\tiny Max} 
    & {\tiny $-v$} \includegraphics[width=0.1\columnwidth]{Figures/isosurface_cmap.png} {\tiny $v$}
    \\
    %\subfigure[]{\includegraphics[width=0.49\columnwidth]{Figures/ball_stick.png}}
    \subfigure[Atom positions ]{\includegraphics[width=0.33\columnwidth]{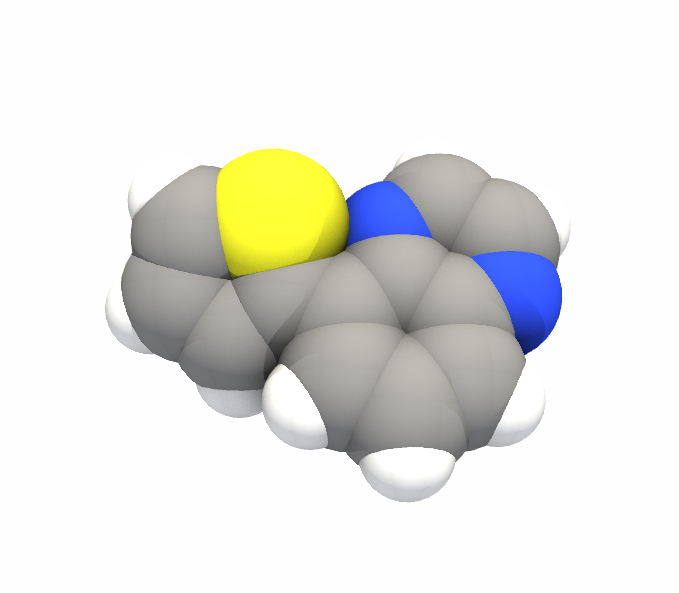}}
    &
    %\subfigure[]{\includegraphics[width=0.49\columnwidth]{Figures/ball_stick_ligands.png}}
    \subfigure[Two subgroups ]{\includegraphics[width=0.33\columnwidth]{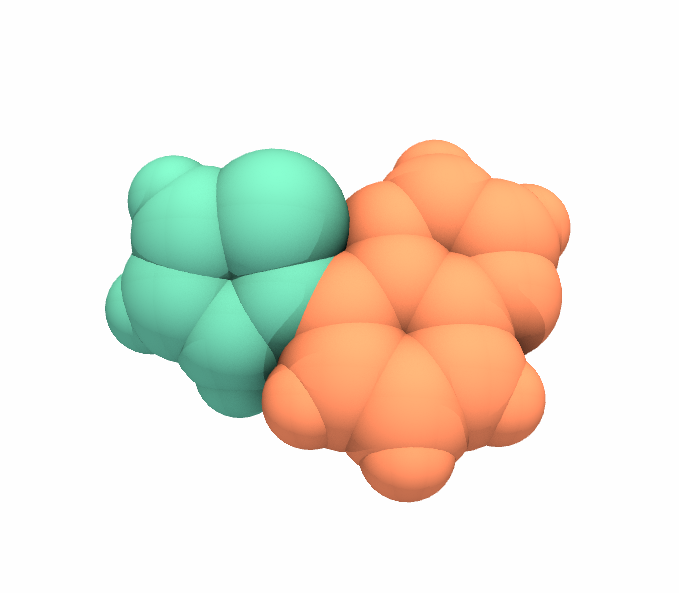}}
    &
    \subfigure[Hole NTO $\Phi_h$]{\includegraphics[width=0.33\columnwidth]{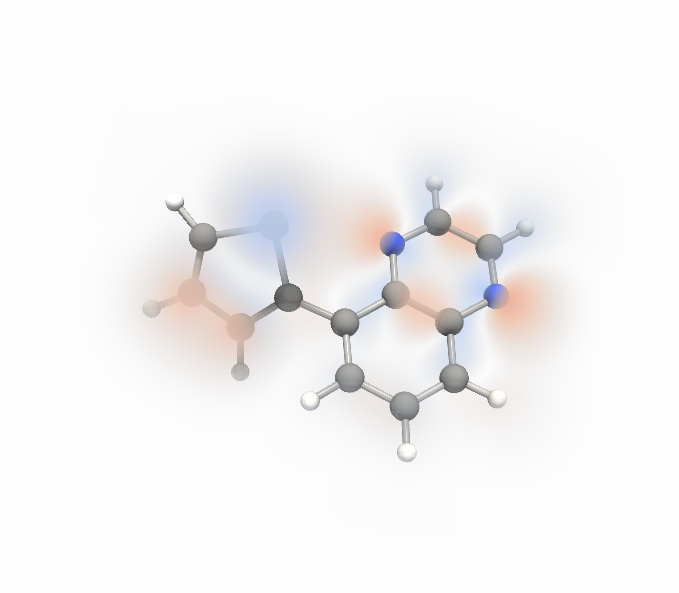}}
    &
    \subfigure[Isosurfaces in $\Phi_h$ ]{\includegraphics[width=0.33\columnwidth]{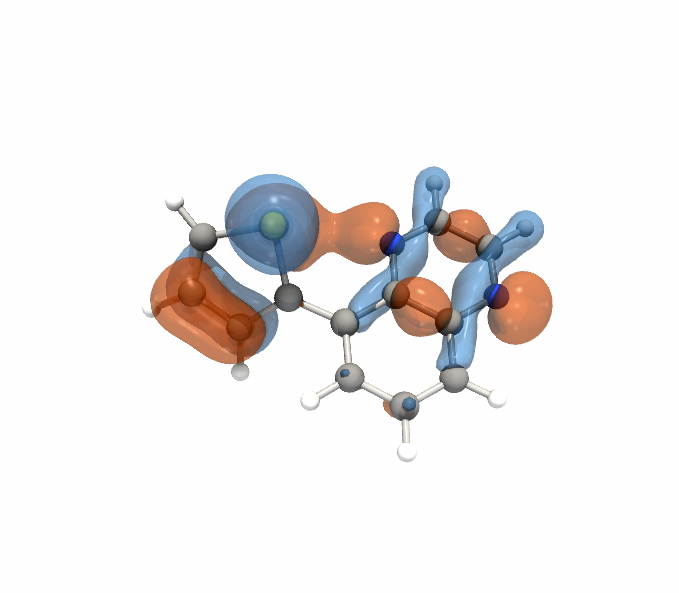}}
    &
    \subfigure[Particle NTO $\Phi_p$]{\includegraphics[width=0.33\columnwidth]{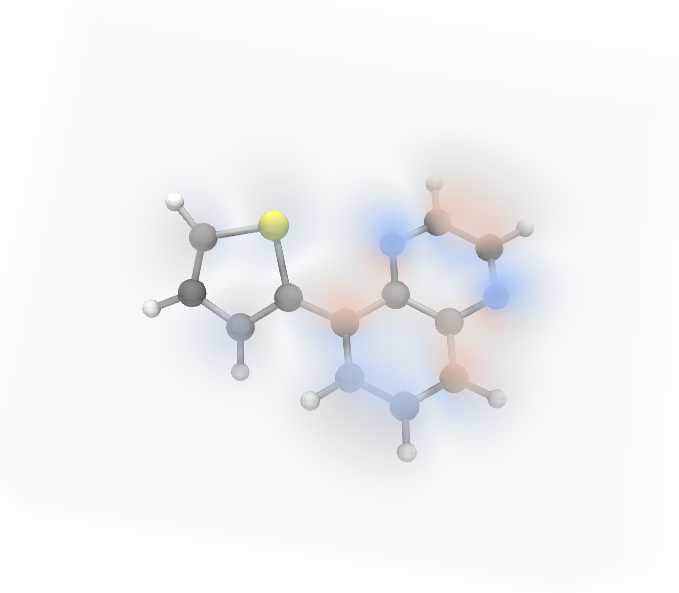}}
    &
    \subfigure[Isosurfaces in $\Phi_p$]{\includegraphics[width=0.33\columnwidth]{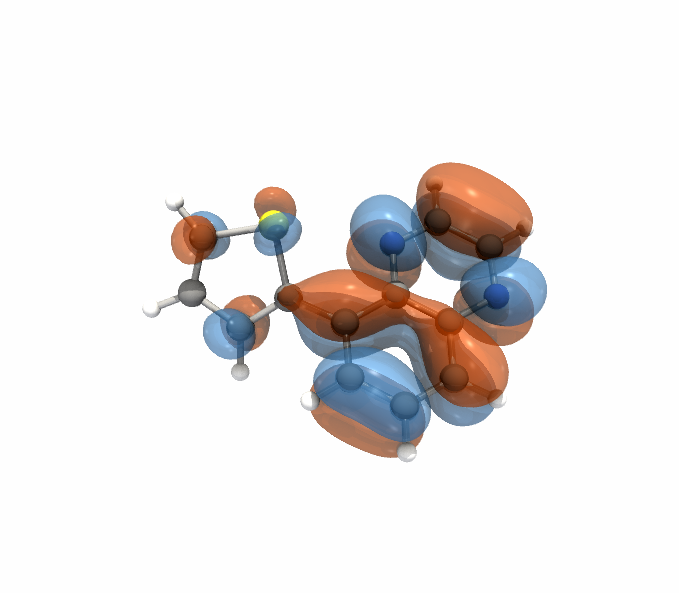}}
    \end{tabular}
    \caption{Direct visualization of the output data from DFT simulations of a Thiophene-Quinoxaline molecule for the first excited state. This data as well as some domain knowledge about potential subgroups serve as input for the automatic analysis and visualization pipeline, Fig.~\ref{fig:overview}.}
    \label{fig:input}
\end{figure*}

The paper is structured as follows. In Sec.~~\ref{section:background}, we summarize required background in chemistry for the application. Sec.~\ref{sec:relatedWork}  summarizes the relevant  related work. The visualization task and analysis problem are formalized in Sec.~\ref{sec:problem}. Sec.~\ref{sec:method} describes in detail the methods of our proposed solution. We present four case studies with increasing complexity in Sec.~\ref{sec:results}. A discussion of the proposed framework 
considering user participation in the visualization design, robustness, and efficiency of the method 
is provided in Sec.~\ref{sec:discussion} before we conclude in Sec.~\ref{sec:conclusion}.

\section{Electronic densities and transitions}
\label{section:background}
Atoms are composed of nuclei and electrons, with the latter occupying most of the space. The occupied space can be interpreted as an electronic cloud or a standing wave called orbital regions with the highest probability to find electrons. Within each atom, there is a series of available orbitals with specific energy levels, each orbital hosting a maximum of two electrons.
In a molecule, the location of electrons is determined by the molecule as a whole, and molecular orbitals are formed by a linear combination of atomic orbitals~\cite{Mulliken}.  Thereby the electrons from the constituent atoms of the molecules fill the molecular orbitals starting from the lowest energy up to the Highest Occupied Molecular Orbital (HOMO) with two electrons per molecular orbital. The remaining orbitals are named Unoccupied orbitals with their lowest called the  Lowest Unoccupied Molecular Orbital (LUMO). 
Molecular Orbitals are used not only to study the electron localization in a molecule but also to calculate chemical and physical properties such as the probability of finding an electron in any specific region or to calculate electronic transitions involved in the interaction between the molecule and light. 
When a molecule absorbs a photon, electrons are excited from the occupied orbitals to unoccupied orbitals.

Mathematically, molecular orbitals are an approximate solution to the Schr\"odinger equation for the electrons in the field of the molecule's atomic nuclei. 
They can be calculated using modern quantum chemistry methods such as Density Functional Theory~(DFT), e.g., implemented in the program Gaussian~\cite{g16}. 
Using the Time-Dependent formalism of DFT (TD-DFT) it is possible to study the electronic transitions within a molecule. The result of such a calculation is excitation energy and a set of coefficients describing the contribution of each orbital to the excited state. An electron promoted from an occupied level will be named a hole on the remaining orbital and will be promoted to a virtual level (particle).
To understand electronic transitions it is not sufficient to look at the individual orbitals but the linear combination of molecular orbitals involved in the electronic transition. Therefore, a more compact orbital representation, named Natural Transition Orbital~(NTO) has been proposed  to describe what has been excited (the hole NTO) and to where it has been excited (the particle NTO)~\cite{NTO}.

The output data of these calculations are scalar values given on a regular grid in a so-called `cube' file.
A typical analysis task is to identify the nature of the electronic transition.
%as a local excitation (LE) or a Charge Transfer excitation (CT).
%
If both the NTO of the particle and hole are located on the same part of the molecule, one speaks of a Local Excitation (LE). In contrast, if the NTO of the particle and hole are located on different parts of the molecule, one speaks of a Charge Transfer excitation (CT). 
This task is generally approached by displaying an isosurface of the resulting scalar fields, Fig.~\ref{fig:input}~(d,f). This is a purely qualitative analysis and there is a need to quantify the character of LE and CT between each relevant part of the molecule. Moreover, a tool to identify quickly the character of each excitation is missing and would be really useful to facilitate the analysis of a series of molecules.  

\section{Related work}
\label{sec:relatedWork}
%{\em Atoms in molecules --} 
\paragraph*{Atoms in molecules.}
Atoms in molecules is a model assuming that the molecular structure can be analyzed using atoms and bonds as constitutive elements. 
The hypothesis is that some characteristic physical properties can be determined on a per-atom basis. 
In this context, the question of how to assign charges to the individual atoms in a molecule is still actively discussed in chemistry with a variety of methods being proposed. Since this segmentation is not a physically observable property but rather a concept supporting the reasoning over molecules, there is no ground truth and the methods have to be validated by their usefulness in a given context.
The methods can be classified into two groups -- wave-function based partitioning in Hilbert space, e.g., Mulliken’s Population analysis~\cite{Mulliken1955}, and 3D space partitioning using an appropriate descriptor, e.g., the electron density distribution of the molecule.  
The electron density distribution is a scalar field, that can be interpreted as the probability of observing the amount of electrons in a specific volume. 
Often the second group is preferred since the results do not depend on the chosen basis functions
for the molecular modeling~\cite{Heidar-Zadeh2018}. 
This second group of methods can be further classified into methods based on direct space partitioning and fuzzy methods following a fractional charge assignment~\cite{Heidar-Zadeh2018}. 
In the case of space partitioning methods, in a second step, the charge density is integrated over pre-determined atomic regions, an area associated with the atom~\cite{Politzer1970}. This approach boils down to the question of how to partition the molecule into atomic regions. For linear molecules, Politzer~\cite{Politzer1970} proposed segmentation of the space using separating planes orthogonal to the molecular axis. This approach has later been generalized to non-linear atoms using a Voronoi partitioning~\cite{Guerra2004}.
While these approaches are purely geometric, a topological field-based segmentation has been proposed by Bader~\cite{Bader1990}.
A fuzzy approach is followed by Hirshfeld~\cite{Hirshfeld1977}, who proposes to share the electronic density at each point among all atoms in relation to their atomic contribution of its spherically averaged ground-state density. 

\paragraph*{Ground and excited state comparison.}
A frequent task is the comparison and characterization of the chemical nature of the electronic ground and excited states. In this context, the comparison of electron charge distributions plays an important role. For this purpose, a set of indexes serving as a  descriptor  for  the  charge  transfer~(CT) have been designed. These indexes are mostly based on a pointwise difference density field which is partitioned in positive and negative regions. The first indexes introduced by Ciofini \etal ~\cite{Garcia2010,Bahers2011} describe CT distance, amount of transferred charge, and the variation of the dipole moment. Later this set has been extended by an index that can be interpreted in terms of hole-electron distance~\cite{Guido2013,Huet2020}. Some of these indexes are also implemented into standard quantum chemical codes, such as Gaussian~\cite{g16}. 
An alternative approach has been proposed by Romouts \etal~\cite{Rombouts2017} who compare the segmented partial charges associated with the individual atoms. They propose to use atom-centered Voronoi cells for the partitioning. In contrast to the analysis of the difference-volume, this approach has the advantage that it is not sensitive to changes in the geometric configuration of the atomic positions. Our approach is closely related to this idea.

\paragraph*{Typical visualization methods used in the domain.}
Visualization plays an important role in most of the above-mentioned studies. The most dominant visualizations are statistical plots and energy diagrams. 
Also, spatial representations are omnipresent. The typically used methods can be summarized as a combination of schematic molecular representations and isosurface plots. 
VMD, a widely used visualization tool for biomolecular systems, even provides hardware support for efficient orbital surface rendering~\cite{Stone2009,Stone2011}.
Isosurfaces, however, can be misleading when the isovalue is not carefully selected and adapted to the orbital energies.
To overcome this limitation, Haranczyk \etal~\cite{Haranczyk2008} propose to visualize
orbitals or electron densities in a more consistent way using a pre-selected fraction of the total charge to determine an orbital-specific isovalue.
A direct visualization of the complex-valued molecular orbitals has been proposed by Al-Saadon \etal~\cite{AlSaadon2019}.
The schematic molecular representations including ball-and-sticks or van der Waals surface representations~\cite{Kozlikova2015} provide an overview of the molecular structures. Electron density isosurface plots provide a view of the density distributions. 
Side-by-side visualization is used for the comparison of ground and excited states. Besides, electron density difference isosurface plots give a more direct impression of the differences in the charge densities. In a few papers, these representations are overlaid with arrows representing the charge transition~\cite{Jacquemin2012}. 
Sometimes, color plots on a slice using a divergent color map can be seen to give a more complete overview of the molecular electrostatic potential ~\cite{Liu2020_b}.  While such plots give a good first qualitative impression about the charge transfer the pictures are not suitable for quantitative analysis. The chosen isovalue is a critical parameter in all these visualizations.

\begin{figure*}[t]
    \centering
    \includegraphics[width=.8\textwidth]{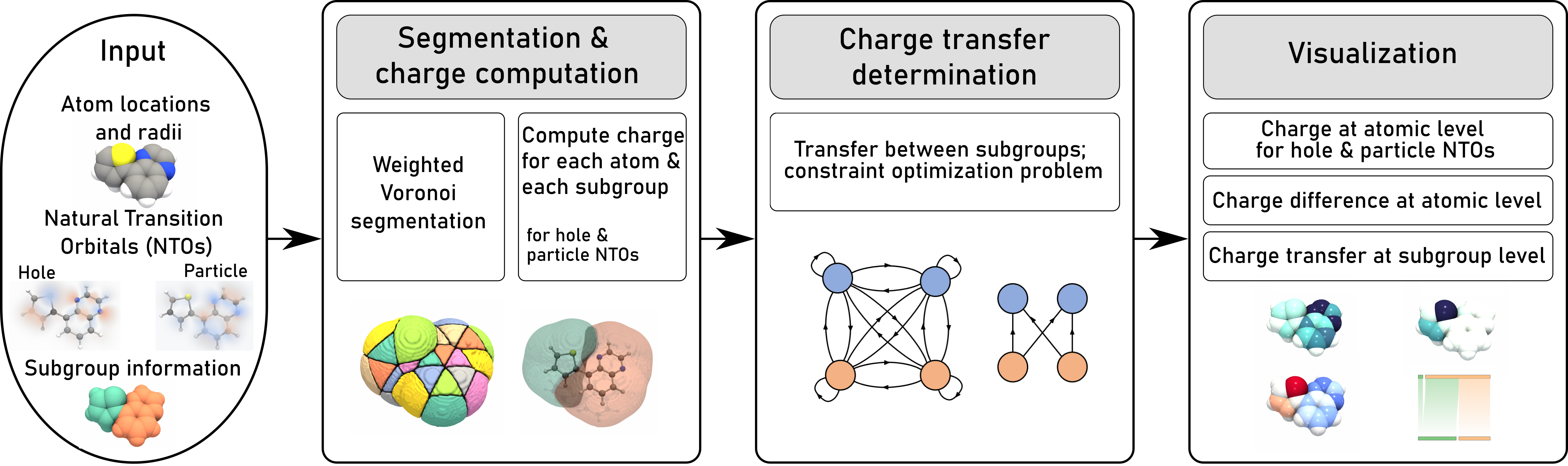}
    \caption{Overview of the method. The first step is to segment the volume and calculate the charge for atoms and subgroups, both for hole and particle Natural Transition Orbitals (NTO), described in Sec.~\ref{section:segmentation}. Then we set up the constraint optimization problem to calculate the charge transfer between subgroups (Sec.~\ref{section:charge_transfer}). The charge results are visualized, both at atomic and subgroup level, this is described in Sec.~ \ref{section:visualization}.}
    \label{fig:overview}
\end{figure*}

\paragraph*{Volume segmentation for visualization.}
Segmentation of scalar fields is a prevalent topic in visualization. Also, the segmentation of the electron density field has been explored to provide insight into the properties of molecules and materials. 
Existing methods can be categorized as geometric and topological methods with two different goals. %and also different requirements. 
The first task is to assign charges to atoms in molecules and it requires high geometric accuracy. %, a task that is similar to ours.
For this purpose, a numerical algorithm to divide the space into regions separated by zero flux surfaces has been presented by Henkelman~\cite{Henkelman2006}. It closely follows the theory introduced by Bader~\cite{Bader1990}. The geometric accuracy of combinatorial approaches is often not sufficient for this purpose. However, there are a few approaches that try to overcome this limitation. Stochastic methods ~\cite{Reininghaus2012a, Gyulassy2012ab} did not provide guarantees but resulted in empirically convergent solutions. 
Facilitating a pre-segmentation, Gyulassy \etal~\cite{Gyulassy2014b} introduced a conforming Morse-Smale complex achieving much better geometric embeddings. Specifically, for the computation of atomic volumes, Bhatia \etal~\cite{Bhatia2018} introduced a computational tool, TopoMS, which combines numerical integration with concepts from computational topology and provides an accurate segmentation while still guaranteeing topological consistency.
The second goal of topological segmentation is the analysis of atomic bonds where geometric accuracy is secondary. However, a robust extraction of the topological skeleton is essential. In this context, methods from computational topology are very successful~\cite{Gunther2014ac}. 
For the analysis of ion diffusion in battery materials, a geometric segmentation of carbon nanospheres, inspired by the Delaunay triangulation, was proposed by Gyulassy \etal~\cite{ Gyulassy2016}. 
Segmentation based on the Voronoi diagram and its dual Delaunay triangulation has also been extensively studied in the context of macromolecules and used to measure their geometric properties such as volume and surface area~\cite{liang1998analytical, petvrek2007mole}.  

\paragraph*{Bipartite graph visualization.}
A common approach to visualize bipartite graphs, which consist of two disjoint sets of nodes, is to represent the two node sets as parallel lines and to draw edges between pairs of nodes that are connected by an edge. For weighted graphs, a frequently used representation is the Sankey diagram encoding the weights or magnitude of the flow in the width of the connecting arrows. 
It was developed over 100 years ago for material flow analysis~\cite{Schmidt2008SankeyDiagram} and is now used in many applications~\cite{Burch2020}. They are especially useful for simple graphs with a limited number of nodes and edges. Some variants can also cope with more complex scenarios including interactive exploration~\cite{Riehmann2005}. 
The chord diagram is another visual representation targeting similar data. Here, the nodes are arranged along a circle and connected by arcs scaled according to the ratio of the out- and in-flow of the respective nodes. 
Such diagrams have been used in many applications including charge flow networks~\cite{Kotravel2019}.
Considering weighted bipartite graphs as a special case of more general flow graphs or transition matrices, all related graph drawing methods are principally applicable~\cite{Nobre2019}. 
There are also methods that focus on the visualization of large scale bipartite graphs with tens of thousands of nodes and edges~\cite{Yeuk-Yin2018} which goes far beyond our needs.

\section{Problem specification and definition}
\label{sec:problem}
In the following, we translate the computational chemists' requirements described above in Sec.~\ref{section:background} into a data-analysis and visualization problem.
We start with specifying the visualization tasks and then give a precise mathematical formulation of the related data analysis problem. 

\subsection{Visualization and analysis tasks}
%Based on the requirements of the domain scientists, 
We identified two data analysis tasks (A1 and A2) and three visualization tasks (V1, V2, and V3) specified below:
\begin{enumerate}
\item[A1] Parameter-free quantification of charges associated to atoms.
\item[A2] Parameter-free quantification of electronic charge transfer between the hole and particle NTO at the chemical subgroup level. 
\item[V1] Visualization of the hole and particle charge distribution.
\item[V2] Visualization of charge differences at atoms to emphasize the loss or gain of electronic charge during the electronic transition. 
\item[V3] Visualization of the charge transfer emphasizing the nature of the electronic transition as Local Excitation (LE) or Charge Transfer excitation (CT). 
\end{enumerate}

The two analysis task relate to quantification of charge and the charge transfer using methods which do not require any data-specific user-defined parameters.  We decided to take a parameter-free approach 
to facilitate easy integration of these methods in automated analysis pipelines for processing a large set of molecules. The three visualization tasks are aimed at discerning the nature of the electronic transition through visual analysis both at the chemical subgroup and atomic level of detail.

\subsection{Formal problem specification} 
The basis for all the visualization tasks is an objective quantification of diverse charge contributions and transitions which should follow clear rules. The data is a set of scalar fields from TD-DFT calculations and the input configuration of the molecules constituted of atoms and atomic groups. The problems to solve are fundamentally partitioning and transfer computation tasks. 

\paragraph*{Input.} We are given the following information: 
\begin{itemize}
\item A set of atoms $A = \{a_1, a_2, \dots, a_N\}$ where each atom $a_i$ is a sphere centered at $p_i=(x_i, y_i, z_i) \in \mathbb{R}^3$ with radius $r_i$, e.g., the van der Waals radius.
\item Partitioning of the atoms into $M$ \emph{subgroups}, $S=\{s_1, s_2, \dots, s_M\}$, where $s_j \subseteq A$, $\cup_{s_j \in S} = A$ and $s_i \cap s_j = \emptyset$ for $i \neq j$.
\item Natural Transition Orbital~(NTO) for the hole $\Phi_h:\mathbb{R}^3 \to \mathbb{R}$ and the particle $\Phi_p:\mathbb{R}^3 \to \mathbb{R}$.
\item In practice the scalar fields $\Phi_h$ and $\Phi_p$ are provided as sampled over a 3D grid $\mathbb{G}$ of size $n_\mathbf{x}\times n_\mathbf{y} \times n_\mathbf{z}$ corresponding to a subset $\mathbb{D} \subset \mathbb{R}^3$. The grid $\mathbb{G}$ consists of voxels of uniform size. 
\end{itemize}

\paragraph*{Problem.} Given this, the problem can be specified as follows:
\begin{itemize}
\item Determine the hole charge $q^h_i$ for each atom $a_i$ such that the total hole charge $\sum_{i=1}^N q^h_i = \int_\mathbb{D} ||\Phi_h||^2$.
\item Similarly, determine the particle charge $q^p_i$ for each atom $a_i$ such that the total particle charge $\sum_{i=1}^N q^p_i = \int_\mathbb{D} ||\Phi_p||^2$.
\item Also, determine the hole charge $Q^h_j$ for each subgroup $s_j$ and the corresponding particle charge $Q^p_j$.
\item Determine the amount of charge transfer $\tilde{Q}_{jk}$ between all pairs of subgroups $s_j$ and $s_k$ according to a few given constraints.
\end{itemize}

\section{Method}
\label{sec:method}

\begin{figure*}[t]
    \centering
    \subfigure[]{\includegraphics[width=0.39\columnwidth]{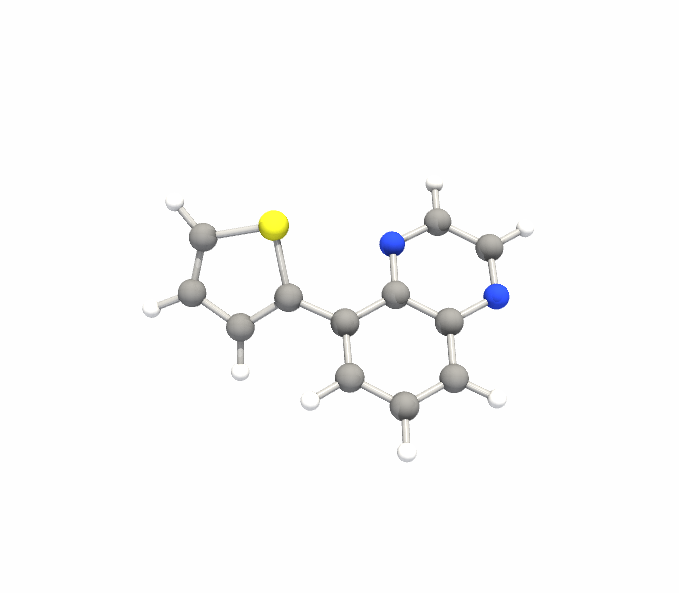} \label{subfig:ballstick_atoms}
    }
    %\subfigure[]{\includegraphics[width=0.4\columnwidth]{Figures/ball_stick_atom_divide.png}}
    \subfigure[]{
    \includegraphics[width=0.39\columnwidth]{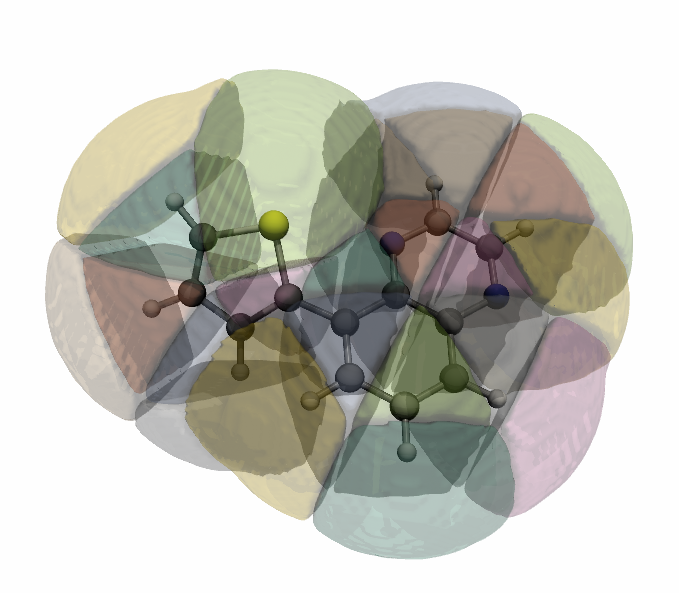} \label{subfig:Voronoi_atoms}
    } 
    \subfigure[]{
    \includegraphics[width=0.39\columnwidth]{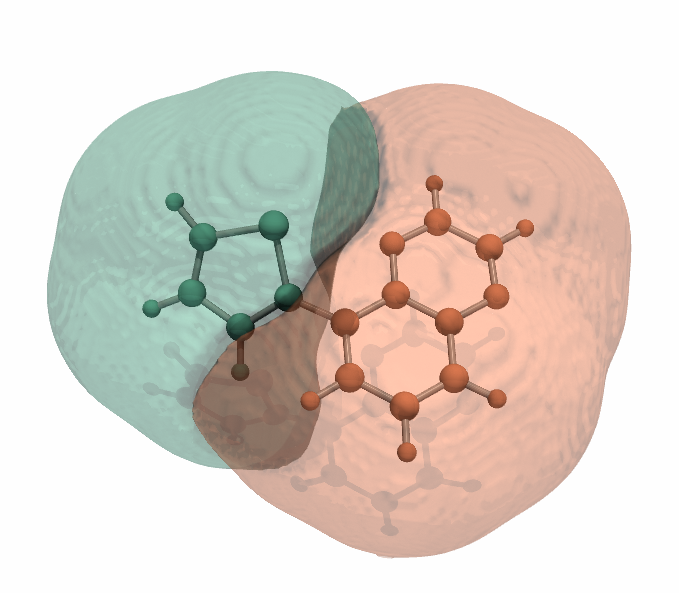} \label{subfig:Voronoi_subgroup}
    } 
    \subfigure[]{
    \includegraphics[width=0.39\columnwidth]{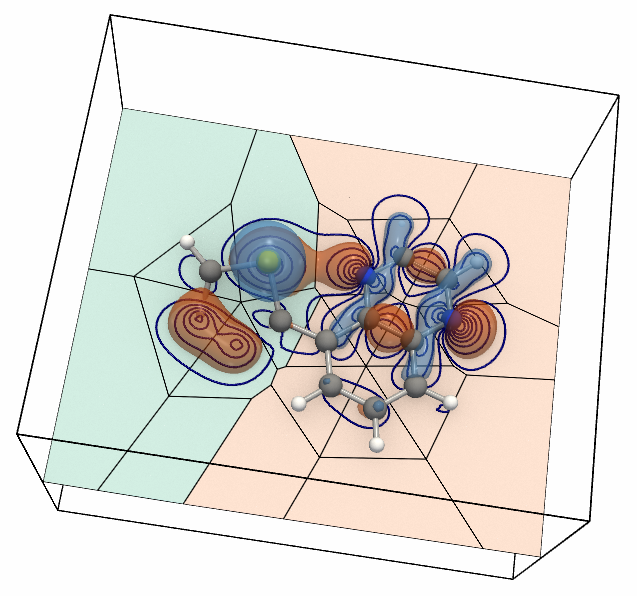} \label{subfig:Voronoi_slice_GS}
    } 
    \subfigure[]{
    \includegraphics[width=0.39\columnwidth]{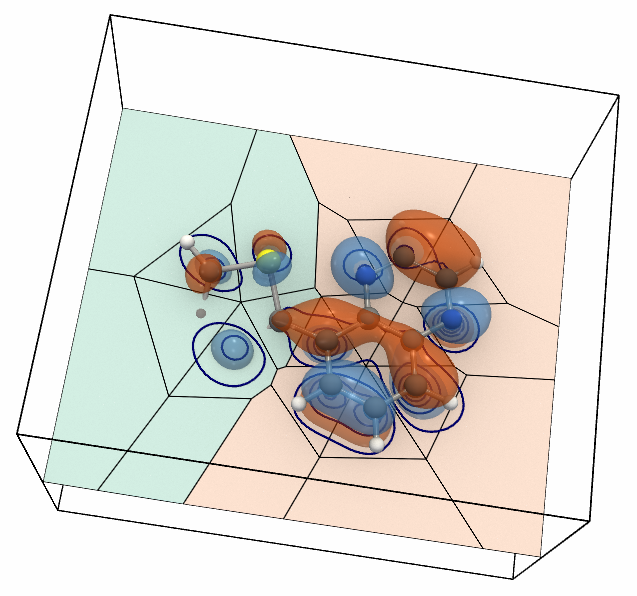} \label{subfig:Voronoi_slice_ES}
    } 
    \caption{Voronoi segmentation for Thiophene-Quinoxaline. \protect\subref{subfig:ballstick_atoms} Ball and stick representation. \protect\subref{subfig:Voronoi_atoms} The weighted Voronoi segmentation. Voronoi segments are clipped by a sphere of radius twice the van der Waals radius for better display. \protect\subref{subfig:Voronoi_subgroup} The combined segments for the two subgroups ({\tiny THIO} \includegraphics[width=0.1\columnwidth]{Figures/subgroup_cmap.png} {\tiny QUIN}). The regions associated to the subgroups are the union of the atomic Voronoi cells. A slice showing the Voronoi segmentation is shown along with isocontours and isosurfaces for the hole NTO $\Phi_h$ \protect\subref{subfig:Voronoi_slice_GS} and the particle NTO $\Phi_p$ \protect\subref{subfig:Voronoi_slice_ES}.} 
    \label{fig:voronoi}
\end{figure*}

An overview of our visual analysis pipeline is presented in Fig.~\ref{fig:overview}. 
The individual steps are described in the following sections.

\subsection{Segmentation and charge computation (Task A1)}
\label{section:segmentation}
To compute the charge of an atom we partition the volume $\mathbb{D}$ into regions, each belonging to one atom. The charge of an atom is then calculated as the charge within the region based on the charge density derived from the NTO, both for particle ($\Phi_p$) and hole ($\Phi_h$).

\paragraph*{Segmentation of the volume.}
We find a partition of the volume $\mathbb{D}$ into non-overlapping regions $V = \{\mathbb{V}_1, \dots, \mathbb{V}_N\}$ such that $\mathbb{V}_i \subseteq \mathbb{D}$ and $\cup_{\mathbb{V}_i \in V} = \mathbb{D}$.
To solve this segmentation problem we considered both topological and geometrical segmentation methods. We investigated the Morse-Smale complex and the weighted Voronoi segmentation and concluded that the resulting subgroup charges were very similar for these two segmentation algorithms, refer to the appendix for a detailed comparison. 
Since we are mainly interested in the subgroup charge and also the fact that we want an efficient and fast method, our choice was the weighted Voronoi due to its simplicity. Chemists are also familiar with this segmentation approach thus making it easier to communicate to domain experts.

\paragraph*{Voronoi segmentation.}
Given $N$ seed points, the Voronoi diagram divides the space into $N$ regions, each region consisting of the points in space closest to one particular seed point \cite{Aurenhammer1991}.
A more general version is the \emph{weighted} Voronoi diagram where each seed point has a weight.
This weighted version is called a \emph{power diagram} when using the power distance to measure the distance  between a point $\mathbf{x}$ and seed point $p_i$ with weight $r_i$: $ pd(\mathbf{x}, p_i) = ||\mathbf{x} - p_i||^2 - r_i^2$ ~\cite{Aurenhammer1987}.
It partitions a volume $\mathbb{D}$ such that the region belonging to atom $a_i$ at position $p_i$ and radius $r_i$ can be described as
\begin{equation}\label{eqn:power_diagram}
    \mathbb{V}_i = \{\mathbf{x} \in \mathbb{D} \text{ such that } pd(\mathbf{x}, p_i) < pd(\mathbf{x}, p_j) \forall j \neq i \}
\end{equation}
This segmentation algorithm only considers the atom positions and their radii together with the volume dimensions and is not dependent on the scalar field values. This means that we do not need to recalculate the segmentation for the different scalar fields $\Phi_h$ and $\Phi_p$ for the same molecule.
An example of a weighted Voronoi segmentation of the volume is shown in Fig.~\ref{fig:voronoi}.

We implement the weighted Voronoi algorithm in 
the discrete setting. In parallel, for each point in the volume grid $\mathbb{G}$, we compare the power distances to all the atom positions and pick the one giving the minimum value. The result will be a volume, where each voxel is labeled with the index of the closest atom. 

\paragraph*{Charge computation in each region.}
The charge density in a point $\mathbf{x}$ is calculated by taking the square of the Natural Transition Orbital value in that point
\begin{equation}\label{eqn:charge_density}
    \rho(\mathbf{x}) = ||\Phi (\mathbf{x})||^2
\end{equation}
Given the set of $N$ regions $V = \{\mathbb{V}_1, \mathbb{V}_2, \dots, \mathbb{V}_N\}$ partitioning our volume $\mathbb{D} \subset \mathbb{R}^3$, and the Natural Transition Orbitals $\Phi_h$ and $\Phi_p$ sampled over $\mathbb{D}$ we want to calculate the sum of all charge densities for each region, the charge for each atom, for both $\Phi_h$ and $\Phi_p$.

We obtain this by using Eqn. \ref{eqn:charge_density} for each point in $\mathbb{V}_i$ and integrating over all points in that region. This integral can be approximated by adding the charge over all voxels $\mathbf{v}_k \in \mathbb{V}_i^\mathbb{G}$ where $\mathbb{V}_i^\mathbb{G} \subset \mathbb{G}$ is the discrete representation of $\mathbb{V}_i$ in $\mathbb{G}$. Let the volume of a voxel in $\mathbb{G}$ be $\mathbf{vol}$, then the charge $q^h_i$ can be computed as:
\begin{equation}\label{eqn:charge_atom_hole}
    q^h_i = \int_{\mathbb{V}_i} \rho(\mathbf{x}) d\mathbf{x} \,\, \simeq \, \mathbf{vol} \sum_{\mathbf{v}_k \in \mathbb{V}_i^\mathbb{G}} ||\Phi_h(\mathbf{v}_k)||^2
    %q_h(a_i) = \sum_{\forall x \in \mathbb{V}_i} ||\Phi_h(x)||^2
\end{equation}
Similarly, the particle charge $q^p_i$ for all the atoms is also computed. 
The charge for each subgroup is the accumulated value for all atoms within the subgroup
\begin{equation}\label{eqn:charge_subgroup_hole}
    Q^h_j  = \sum_{a_i \in s_j} q^h_i \,, \qquad Q^p_j  = \sum_{a_i \in s_j} q^p_i
\end{equation}

\subsection{Charge transfer (Task A2)}
\label{section:charge_transfer}
To study the charge transfer at the level of subgroups, it would be beneficial to quantify the charge transfer between two subgroups. However, there is no unique solution for this problem as the system of equations is underdetermined. We describe this in the following subsections and propose some solutions for the charge transfer problem under reasonable assumptions. 

\paragraph*{Problem specification.}
The charge transfer problem requires determination of the charge transfer $\tilde{Q}_{ij}$ from a subgroup $s_i$ to another subgroup $s_j$ under two constraints: (1)~the total charge transfer from a subgroup $s_i$ should be equal to the hole charge $Q^h_i$ of the subgroup, and (2)~the total charge transfer to a subgroup $s_j$ should be equal to the particle charge $Q^p_j$ of the subgroup. This problem can be written in matrix form as follows.
\begin{align}
    \text{Determine }\; & \tilde{\mathcal{Q}}_{M\times M} = 
    \begin{bmatrix}
        \tilde{Q}_{11} & \dots & \tilde{Q}_{1M}\\
        \vdots & \ddots & \vdots\\
        \tilde{Q}_{M1} & \dots & \tilde{Q}_{MM}\\
    \end{bmatrix} \label{eqn:matrix_CT}
    \\
    \text{Such that }\;\; & 
    \sum_{j=1}^M \tilde{Q}_{ij} = Q^h_i \;\;\text{ and }%\label{eqn:row_constraint_CT}
    \;\;%\\ &
    \sum_{i=1}^M \tilde{Q}_{ij} = Q^p_j
    \label{eqn:constraints_CT}
\end{align}

\begin{figure}[t!]
    \centering
    \includegraphics[width=\columnwidth]{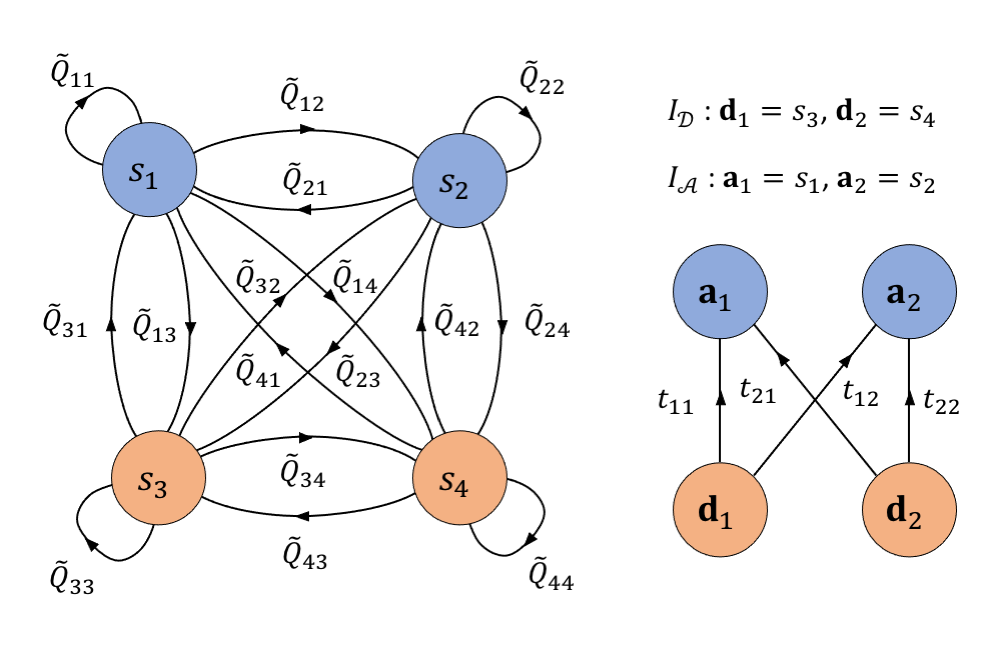}
    \caption{The general charge transfer problem for four subgroups (left) and a simpler formulation (right).}
    \label{fig:graph_rep}
\end{figure}

The matrix $\tilde{\mathcal{Q}}_{M\times M}$ can be interpreted as a weighted complete directed graph. The subgroups $S$ correspond to the vertex set and the matrix elements correspond to the edge weights. See Fig.~\ref{fig:graph_rep}~(left) for an example with four subgroups.

\paragraph*{A simpler problem.} 
A sub-group $s_i$ is called a \emph{donor} if $Q^h_i>Q^p_i$. Otherwise, it is called an \emph{acceptor}. 
We define the charge difference for $s_i$ as $Q^d_i=Q^p_i -Q^h_i$.
Clearly the set S can be partitioned into two subsets, one consisting of donors $\mathcal{D}=\{\mathbf{d}_1,\mathbf{d}_2,\dots,\mathbf{d}_n\}$ and the other consisting of acceptors $\mathcal{A}=\{\mathbf{a}_1,\mathbf{a}_2,\dots,\mathbf{a}_m\}$. Here, $n=|\mathcal{D}|$ is the number of donors and $m=|\mathcal{A}|$ is the number of acceptors such that $n+m=M$. Let $I_\mathcal{D}:\{1, \dots, n\}\to\{1, \dots, M\}$ be the index map that maps a donor $\mathbf{d}_i, i \in \{1, \dots, n\}$ to the corresponding subgroup $s_j, j \in \{1, \dots, M\}$. Similarly, let  $I_\mathcal{A}:\{1, \dots, m\}\to\{1, \dots, M\}$ be the index map mapping the acceptors to the corresponding subgroup $s_j$ in the original list of subgroups $S$.

Now, we make the assumption that there is no charge transfer from an acceptor to any other subgroup, and no charge transfer from a donor to other donors. This implies for all $i \neq j \in \{1, \dots, M\}$
\begin{align}
     & \tilde{Q}_{ij} = 0 \text{ if } s_i \in \mathcal{A},  \label{eqn:acceptor_zero_transfer}\\
\text{and } & \tilde{Q}_{ij} = 0 \text{ if } s_i, s_j \in \mathcal{D}. \label{eqn:donor_zero_transfer}
\end{align}
This assumption, along with the two constraints in~\ref{eqn:constraints_CT}, lead to the following assignment of values to the diagonal of the matrix $\tilde{\mathcal{Q}}$:
\begin{align}
     & \tilde{Q}_{ii} = Q^h_i \text{ if } s_i \in \mathcal{A}, \label{eqn:acceptor_self_transfer}\\
\text{and } & \tilde{Q}_{ii} = Q^p_i \text{ if } s_i \in \mathcal{D}. \label{eqn:donor_self_transfer}
\end{align}

Considering equations~\ref{eqn:acceptor_zero_transfer} to \ref{eqn:donor_self_transfer}, we can remove all the diagonal elements from $\tilde{\mathcal{Q}}_{M\times M}$ and the non-diagonal elements which are $0$ to obtain a smaller matrix $T_{n\times m}$ with $n$ rows corresponding to the donors and $m$ columns corresponding to the acceptors.
The matrix $T$ represents weights of the edges in a complete bipartite graph with directed edges from the set $\mathcal{D}$ to $\mathcal{A}$. 
See Fig.~\ref{fig:graph_rep} for an example.
$t_{ij}$ denotes the charge transfer between donor $\mathbf{d}_i$ and acceptor $\mathbf{a}_j$. The charge transfer for all donor-acceptor pairs is represented as
\begin{equation*}\label{eqn:matrix_T}
    T_{n\times m} = 
    \begin{bmatrix}
        t_{11} & \dots & t_{1m}\\
        \vdots & \ddots & \vdots\\
        t_{n1} & \dots & t_{nm}\\
    \end{bmatrix}
\end{equation*}
%
%Note that we can use 
The index maps $I_\mathcal{D}$ and $I_\mathcal{A}$ provide the element corresponding to $t_{ij} \in T_{n\times m}$ within the matrix $\tilde{\mathcal{Q}}_{N\times N}$ as $t_{ij}= \tilde{Q}_{I_\mathcal{D}(i)I_\mathcal{A}(j)}$. We use the notation $Q_{\mathbf{d_i}}$ to mean $Q_{I_\mathcal{D}(i)}$ and $Q_{\mathbf{a_j}}$ to mean $Q_{I_\mathcal{A}(j)}$

The total charge transfer from a donor $\mathbf{d}_i$ to acceptors should be equal to the charge deficit at $\mathbf{d}_i$. So, for all $\mathbf{d}_i\in D$,
\begin{equation}\label{eqn:rowsum}
    \sum_{j=1}^m t_{ij} = Q^h_{\mathbf{d_i}}-Q^p_{\mathbf{d_i}}=-Q^d_{\mathbf{d_i}}
\end{equation}
The above set of equations gives us $n$ \emph{row-sum} constraints. Similarly, total charge transfer to an acceptor $\mathbf{a_j}$ from all donors should be equal to the charge excess at $\mathbf{a_j}$. So, for all $\mathbf{a_j}\in \mathcal{A}$:
\begin{equation}\label{eqn:colsum}
    \sum_{i=1}^n t_{ij} = Q^p_{\mathbf{a_j}}-Q^h_{\mathbf{a_j}}=Q^d_{\mathbf{a_j}}
\end{equation}
This provides the additional $m$ \emph{column-sum} constraints. The constraints~\ref{eqn:rowsum} and \ref{eqn:colsum} follow directly from \ref{eqn:constraints_CT} respectively. The goal is to determine matrix $T$ under these two constraints~\ref{eqn:rowsum} and \ref{eqn:colsum}. Additionally, we have the \emph{non-negativity} constraints,
\begin{equation}\label{eqn:non_negative}
    t_{ij} \geq 0
\end{equation}
In total, we have $n+m$ equations, of which $n+m-1$ are linearly independent. However, we have $n\times m$ unknowns. A unique solution exists only in the scenario when there is only one donor or in the case when there is only one acceptor.

\paragraph*{Proportional charge division approach.}
One way to arrive at a solution for the charge transfer problem would be to assume $t_{ij}$ to be directly \emph{proportional} to the charge differences at the donor $\mathbf{d_i}$ and the acceptor $\mathbf{a_j}$. That is, $t_{ij} \propto Q^d_{\mathbf{a_j}}\times Q^d_{\mathbf{d_i}}$.
With this assumption and the constraints \ref{eqn:rowsum} and \ref{eqn:colsum}, we arrive at a unique solution for $t_{ij}$ which we refer to as \emph{proportional} solution:
\begin{equation}
    %t_{ij} = \frac{-Q^d_{\mathbf{a_j}}\times Q^d_{\mathbf{d_i}}}{\tilde{Q}},
    t_{ij} = -Q^d_{\mathbf{a_j}}\times Q^d_{\mathbf{d_i}}/\tilde{Q},
\end{equation}
where $\tilde{Q}$ is the total charge transfer: 
\begin{equation}\label{eqn:defineC}
    \tilde{Q} = \sum_{i=1}^n \sum_{j=1}^m t_{ij} = \sum_{i=1}^n -Q^d_{\mathbf{d_i}} = \sum_{j=1}^m Q^d_{\mathbf{a_j}}
\end{equation}

It is easy to observe that such a solution satisfies the \emph{non-negativity} constraint~(\ref{eqn:non_negative}). Further, we show that it also satisfies the \emph{row sum} constraint~(\ref{eqn:rowsum}) and  \emph{column sum} constraint~(\ref{eqn:colsum}):
\begin{align*}
    \sum_{j=1}^m t_{ij} &= \sum_{j=1}^m \frac{-Q^d_{\mathbf{a_j}}\times Q^d_{\mathbf{d_i}}}{\tilde{Q}}
    &= \frac{-Q^d_{\mathbf{d_i}}}{\tilde{Q}}  \sum_{j=1}^m Q^d_{\mathbf{a_j}}
    &= \frac{-Q^d_{\mathbf{d_i}}}{\tilde{Q}} \cdot \tilde{Q} = -Q^d_{\mathbf{d_i}}
    \\
    \sum_{i=1}^n t_{ij} &= \sum_{i=1}^n \frac{-Q^d_{\mathbf{a_j}}\times Q^d_{\mathbf{d_i}}}{\tilde{Q}} 
    &= \frac{Q^d_{\mathbf{a_j}}}{\tilde{Q}}  \sum_{i=1}^n -Q^d_{\mathbf{d_i}}
    &= \frac{Q^d_{\mathbf{a_j}}}{\tilde{Q}} \cdot \tilde{Q} = Q^d_{\mathbf{a_j}}
\end{align*}

\begin{figure*}[!tb]
    \centering
    \begin{tabular}{c@{\hskip1pt}c@{\hskip1pt}c@{\hskip1pt}c@{\hskip1pt}c@{\hskip1pt}c@{\hskip1pt}c}
    \toprule
    & \textbf{$\Phi_h$:} {\tiny$-v$} \includegraphics[width=0.12\columnwidth]{Figures/isosurface_cmap.png} {\tiny $v$}   
    & \textbf{$\Phi_p$:} {\tiny $-v$} \includegraphics[width=0.12\columnwidth]{Figures/isosurface_cmap.png} {\tiny $v$}
    & \textbf{$q^h$:} \includegraphics[width=0.12\columnwidth]{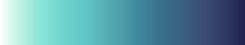} 
    & \textbf{$q^p$:} \includegraphics[width=0.12\columnwidth]{Figures/blue_linear.png} 
    & \textbf{$q^d$:} \includegraphics[width=0.12\columnwidth]{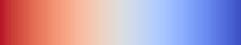} 
    & \textbf{Transition} 
    \\ 
    \midrule
    \raisebox{0.9\height}{\rotatebox{90}{\textbf{State 4}}}
    &
    %\subfigure[]{
    \includegraphics[width=0.28\columnwidth]{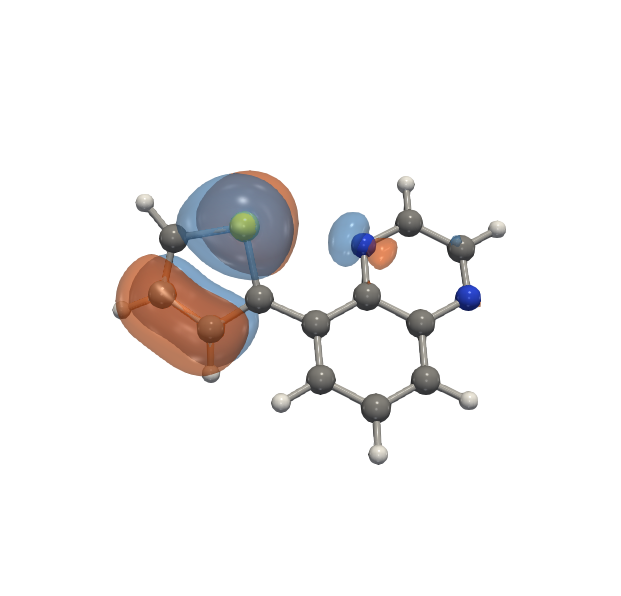}
    %}
    &
    %\subfigure[]{
    \includegraphics[width=0.28\columnwidth]{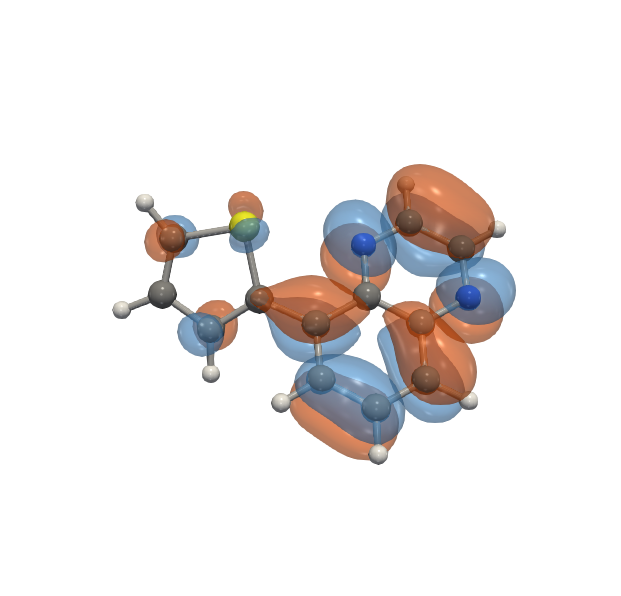}
    %}
    &
    %\subfigure[]{
    \includegraphics[width=0.29\columnwidth]{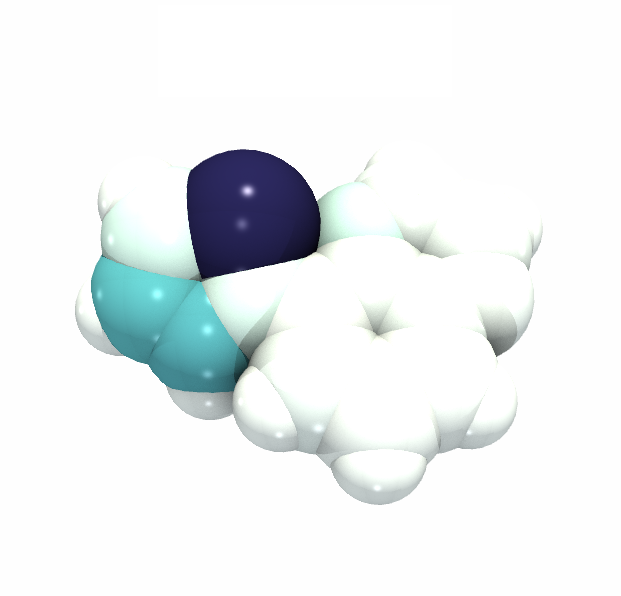}
    %}
    &
    %\subfigure[]{
    \includegraphics[width=0.29\columnwidth]{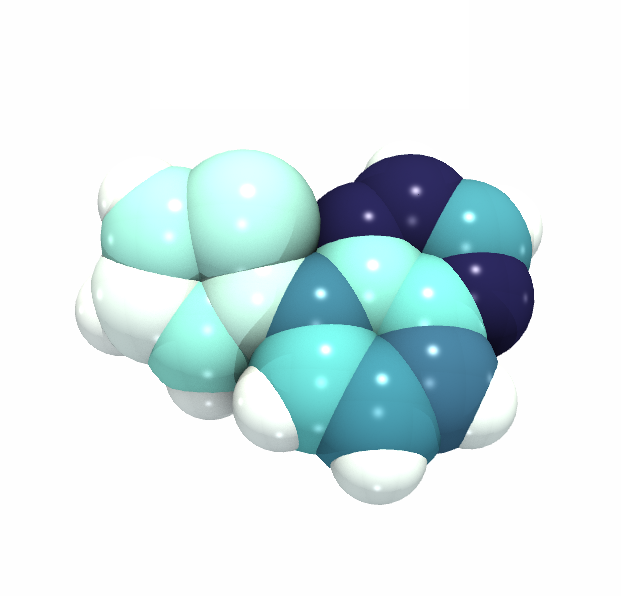}
    %}
    &
    %\subfigure[]{
    \includegraphics[width=0.29\columnwidth]{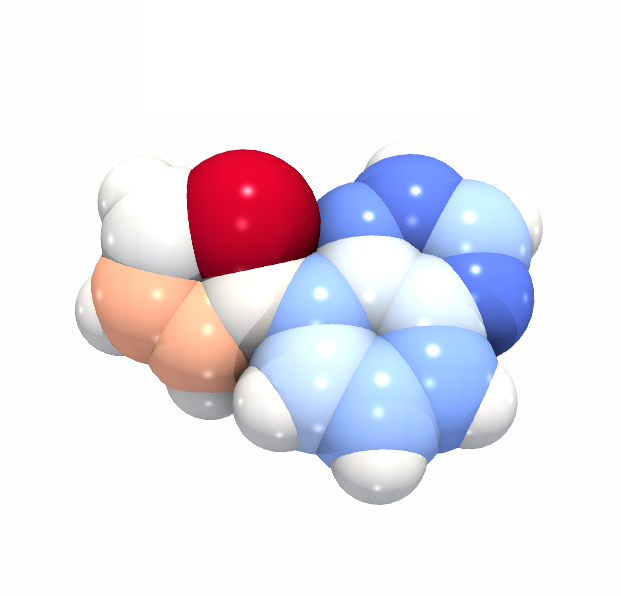}
    %}
    &
    %\subfigure[]{
    \includegraphics[height=0.29\columnwidth]{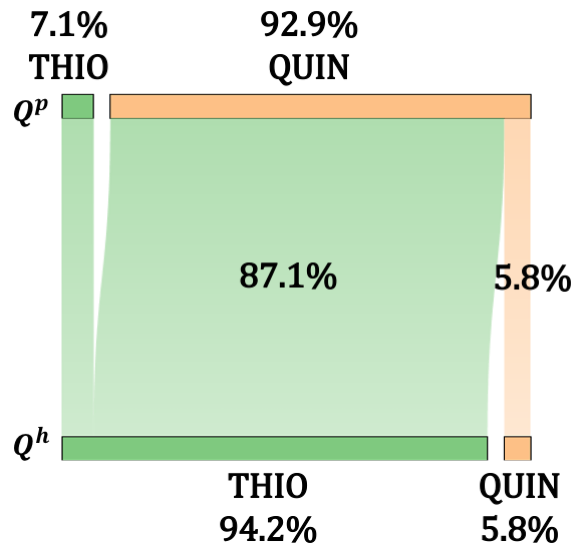}
    %}
    \\
    \raisebox{0.9\height}{\rotatebox{90}{\textbf{State 9}}}
    &
    %\subfigure[]{
    \includegraphics[width=0.28\columnwidth]{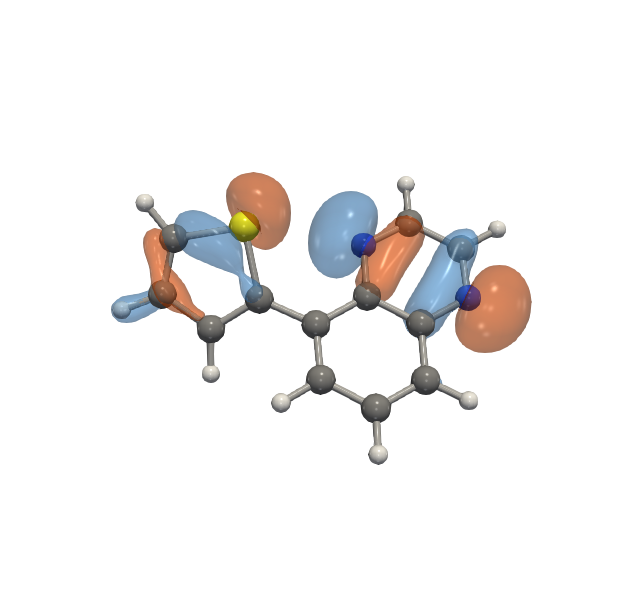}
    %}
    &
    %\subfigure[]{
    \includegraphics[width=0.28\columnwidth]{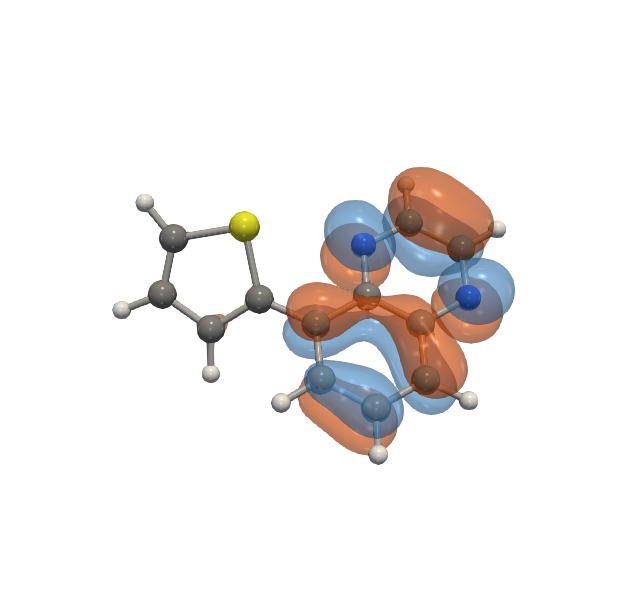}
    %}
    &
    %\subfigure[]{
    \includegraphics[width=0.29\columnwidth]{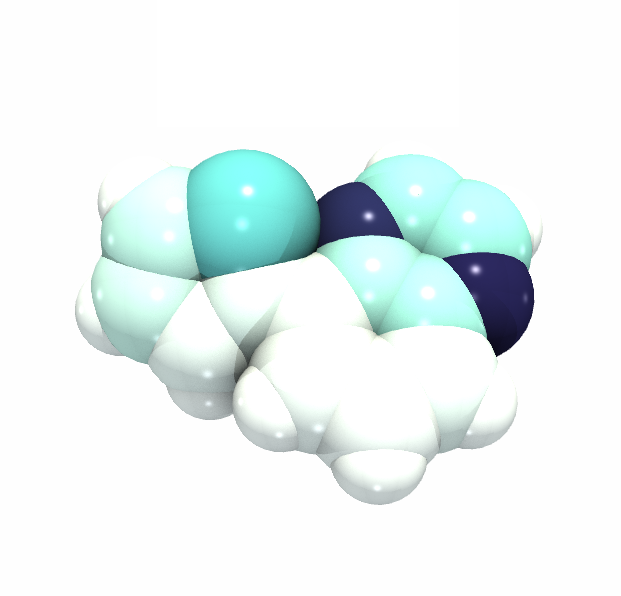}
    %}
    &
    %\subfigure[]{
    \includegraphics[width=0.29\columnwidth]{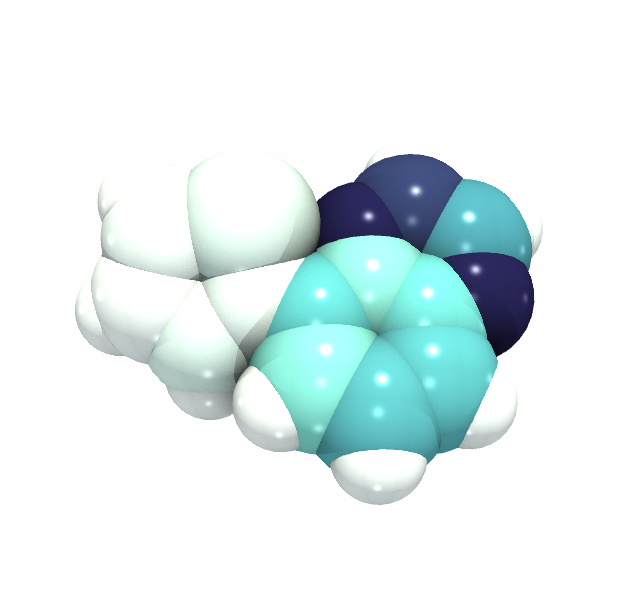}
    %}
    &
    %\subfigure[]{
    \includegraphics[width=0.29\columnwidth]{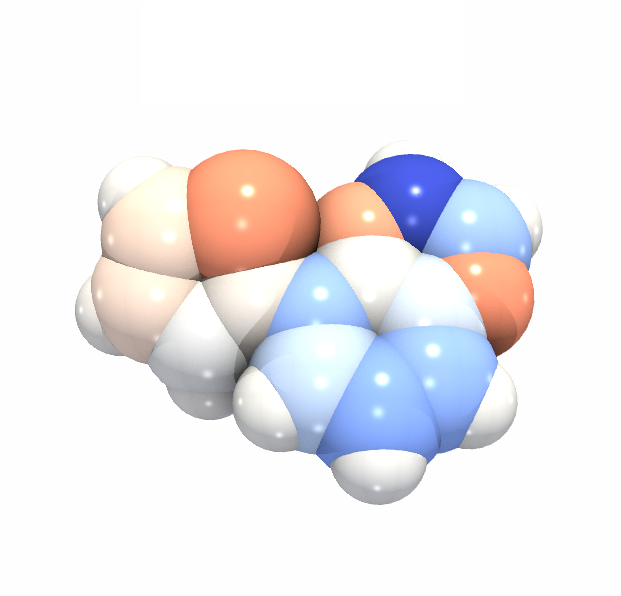}
    %}
    &
    %\subfigure[]{
    \includegraphics[height=0.29\columnwidth]{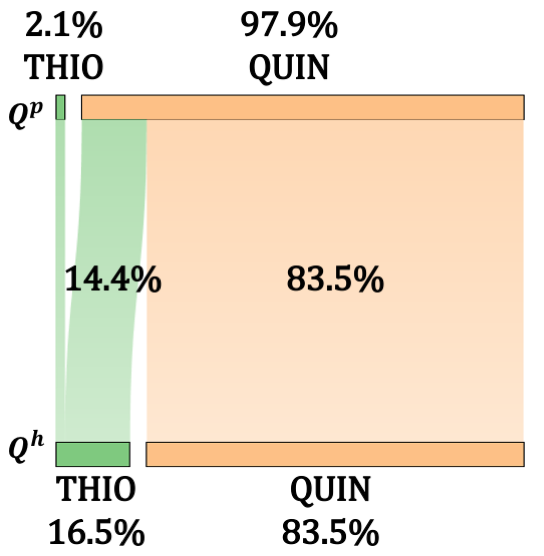}
    %}
    \end{tabular}
    \caption{
    Two different state transitions in the Thiophene-Quinoxaline molecule. Top row: 
    \emph{charge transfer transition}. Bottom row: 
    \emph{local excitation}. 
    Cols. 1 and 2 relate to visualization task V1, cols. 3-5 to the task V2, and the col. 6 to the task V3.
    }
    \label{fig:tq-caseStudy}
\end{figure*}

\paragraph*{Quadratic optimization approach.} We now discuss a  general approach that provides the flexibility of determining an optimal solution under different assumptions and criteria. The matrix $T_{n\times m}$ can be represented as a column vector $\mathbf{t}$ having $n\times m$ components by sequencing the terms $t_{ij}$ in row major order. Constraints in \ref{eqn:rowsum} and \ref{eqn:colsum} can be combined as an \emph{equality constraint}
\begin{equation}\label{eqn:equalityConstraint}
    B \mathbf{t} = \mathbf{b},
\end{equation}
where $B$ is an $(n+m-1) \times (n\times m)$ matrix containing the binary coefficients of $n+m-1$ linearly independent constraints as the rows. 
The $n$ \emph{row constraints} are specified first followed by the $m-1$ \emph{column constraints}. The vector $\mathbf{b} = [-Q^d_{\mathbf{d_1}}, \dots, -Q^d_{\mathbf{d_n}}, Q^d_{\mathbf{a_1}}, \dots, Q^d_{\mathbf{a_{m-1}}}]^T$ corresponds to the value for these constraints. 
Secondly, we have the \emph{non-negativity constraint} as specified in \ref{eqn:non_negative} that is converted into the \emph{inequality constraint}:
\begin{equation}\label{eqn:inequalConstraints}
    \mathbf{t} \geq 0
\end{equation}

Let us consider the simple case of two donors and two acceptors to understand this conversion to vectors. The matrix $T$ is:
\[
    T_{2\times2} = 
    \begin{bmatrix}
        t_{11} & t_{12}\\
        t_{21} & t_{22}\\
    \end{bmatrix}
\]
The four constraints are:
\begin{align*}
    t_{11} + t_{12} & = -Q^d_{\mathbf{d_1}} &\quad 
    t_{21} + t_{22} & = -Q^d_{\mathbf{d_2}} \\
    t_{11} + t_{21} & = Q^d_{\mathbf{a_1}} &\quad 
    t_{12} + t_{22} & = Q^d_{\mathbf{a_2}}
\end{align*}
We can ignore one of the above constraints resulting in a set of linearly independent constraints. Let us assume we ignore the last equation.
Then vectors $\mathbf{t}=[t_{11}, t_{12}, t_{21}, t_{22}]^T$ and $\mathbf{b}=[-Q^d_{\mathbf{d_1}}, -Q^d_{\mathbf{d_2}}, Q^d_{\mathbf{a_1}}]^T$ while the constraint matrix $B$ is:
\[
    B = 
    \begin{bmatrix}
        1 & 1 & 0 & 0 \\
        0 & 0 & 1 & 1 \\
        1 & 0 & 1 & 0 \\
    \end{bmatrix}
\]

It is reasonable to assume that there is a uniform transfer of charge from any donor to any acceptor without any preference. Under this assumption the preferred transfer vector would be $\mathbf{t}_p = [\tilde{Q}/(n\times m), \dots, \tilde{Q}/(n\times m)]$ where $\tilde{Q}$ is as defined in \ref{eqn:defineC}. However, $\mathbf{t}_p$ may not satisfy the constraints in Equation~\ref{eqn:equalityConstraint}. In this case, we can set up an optimization problem to find the optimal $\mathbf{t}_o$ which is closest to $\mathbf{t}_p$ but satisfies the constraints \ref{eqn:equalityConstraint} and \ref{eqn:inequalConstraints}. More formally, we can write this as the following optimization problem:
\begin{equation}
  \label{eqn:optimize}
  \begin{aligned}
    \text{Minimize} \qquad & ||\mathbf{t} - \mathbf{t}_p||^2 \\
    \text{subject to} \qquad & \mathbf{t} \geq 0 \quad \text{and} \quad B \mathbf{t} = \mathbf{b} 
  \end{aligned}
\end{equation}

We recognize this as a quadratic optimization problem that can be solved to obtain an optimal solution $\mathbf{t}_o$ which we refer to as \emph{quadratic solution}. Note that $\mathbf{t}_p$ does not have to be a uniform vector, it can have different weights and allows the possibility of computing optimal $\mathbf{t}_o$ under different scenarios. 

\subsection{Visualization}
\label{section:visualization}
We use various spatial and information visualization methods to facilitate detailed visual analysis of electronic transitions. We describe these methods in brief in the following paragraphs. The implementation was primarily done in Python using VTK~\cite{schroeder2004visualization}. We also used Inviwo~\cite{jonsson2019inviwo} and Paraview~\cite{ahrens2005paraview} for prototyping and generating some higher quality images.

\paragraph*{Spatial visualization (Task V1, V2).} We employ standard scalar field visualization techniques like volume rendering, slicing and isosurface extraction for visualization of the orbitals $\Phi_h$, $\Phi_p$ and the derived density fields $\rho_h = ||\Phi_h||^2$ and $\rho_p = ||\Phi_p||^2$. For direct volume rendering of scalar field $\Phi_h$ and $\Phi_p$, we employ a diverging blue to red color map with a V-shaped transfer function for opacity so that the voxels with extreme positive and negative values are emphasized, see Fig.~\ref{fig:input}~(c,e). 
Two isosurfaces represent the orbitals $\Phi_h$ and $\Phi_p$, for a given value $v$, one for value $v$ and the other for $-v$. 
They are displayed together colored orange and blue respectively, see the first two columns of Fig.~\ref{fig:tq-caseStudy}. 
For visualization of the fields $\rho_h$ and $\rho_p$, we use a linear color map and opacity transfer functions. 

The molecules are shown embedded in the same volume represented either as ball-and-stick or the space-fill model with the radii scaled proportionally to their van der Waals radii. We use color mapping on the balls representing the atoms to show various scalar quantities like hole charge $q^h$, particle charge $q^p$ and charge difference $q^d$ defined as $q^p-q^h$. In case of charge difference, we use a red-to-blue diverging color map highlighting the donor atoms in shades of red and acceptor atoms in shades of blue, see the $q^d$ column of Fig.~\ref{fig:tq-caseStudy}. 
For $q^h$ and $q^p$, a linear white-to-blue color map is used, refer to the $q^h$ and $q^p$ columns in Fig.~\ref{fig:tq-caseStudy}. Additionally, we support the coloring of atoms based on atomic type using the Corey-Pauling-Koltun~(CPK) color model. Lastly, color mapping to highlight the different subgroups is also supported, see Fig.~\ref{fig:input}(a, b).

Within this 3D spatial visualization, we also support the visualization of the Voronoi segments corresponding to the atoms or the combined Voronoi segments corresponding to the subgroups, see Fig.~\ref{fig:voronoi}. 
We do not use direct volume rendering for such a visualization as it suffers from occlusion and discretization artifacts. Instead, we extract the bounding surface of each segment separately and smoothen the surface using a standard Gaussian filter. Further, the segment for atom $a_i$ is clipped by a ball of radius $2r_i$. 

\paragraph*{2D visualization (Task V3).} To further support visual analysis, especially at the subgroup level, we utilize standard information visualization methods. The charges $Q^h$ and $Q^p$ on various subgroups are plotted as bar charts. To visualize the charge transfer we use a \emph{transition diagram}. It is a variant of parallel set visualization~\cite{Kosara2006ParallelSets} and Sankey diagram~\cite{Schmidt2008SankeyDiagram} which we have modified for this particular application,
last column of the Fig.~\ref{fig:tq-caseStudy}. At the bottom of the transition diagram, a bar of width proportional to the hole charge $Q^h$ is displayed for each subgroup in the molecule. At the top, the width of the bars is proportional to the particle charge $Q^p$. In the middle, connectors 
are drawn such that their width is proportional to the charge transfer $\tilde{Q}_{ij}$. The colors of the bars are based on the subgroups and can be chosen by the user. Brushing and linking is supported between the 3D spatial view and the 2D visualizations.
 
\section{Results}
\label{sec:results}
The data for the case studies is calculated using the Gaussian software package~\cite{g16}. The cube files for individual molecular orbitals were generated using the included gencube program. The particle and hole NTOs were generated
based on the set of coefficients describing the contribution that each particle-hole pair makes to the excited state. 

\subsection{Case study 1: Thiophene-Quinoxaline}
As a first simple example, we study the thiophene-quinoxaline molecule which is composed of two subgroups as shown in Figure~\ref{fig:input}(b).
Polythiophene polymer has been widely used in the field of organic electronics and in particular in organic field-effect transistors and organic solar cells because of its high conductivity. %(up to 1 S/cm). 
Thiophene and quinoxaline have very different properties since they are an acceptor and a donor of electrons, respectively. 
These properties provoke a charge transfer from the thiophene to the quinoxaline moieties making this a perfect model system to test our analysis and visualization pipeline. 
As can be seen in Fig.~\ref{fig:tq-caseStudy}, the 4th and 9th excited states have very different properties. For excited state 4, the hole is localized on the thiophene moiety and the particle on the quinoxaline moiety.
However, for the excited state 9, the hole is more delocalized over the whole molecule. This difference can be observed qualitatively by looking at the charge distribution for the hole and the particle as well as the charge difference between them. All this information is included in the transition diagram that provides a quantitative measurement of the nature of the transition. Here, a charge transfer of 87.1\% between the thiophene and quinoxaline for state 4, and a local excitation of the quinoxaline (83.5\%) for state 9 is observed.

\subsection{Case study 2: [6]cycloparaphenylene}
\begin{figure}[b]
    \centering
    \subfigure[Atoms]{\includegraphics[width=0.32\columnwidth]{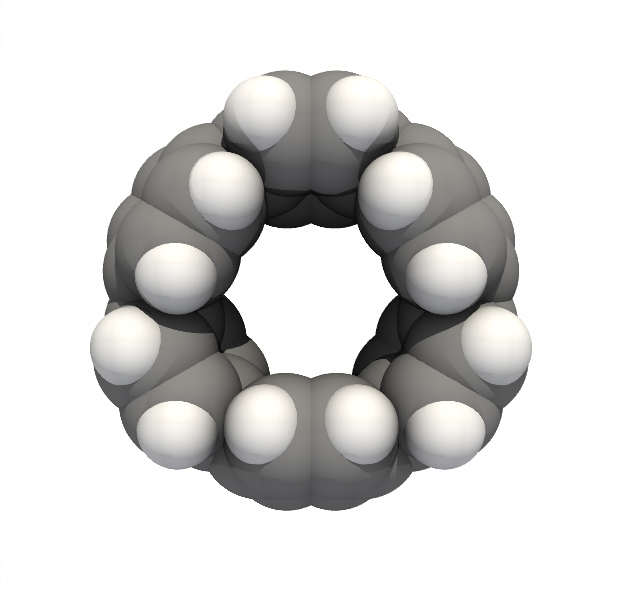}}
    \subfigure[Subgroups]{\includegraphics[width=0.32\columnwidth]{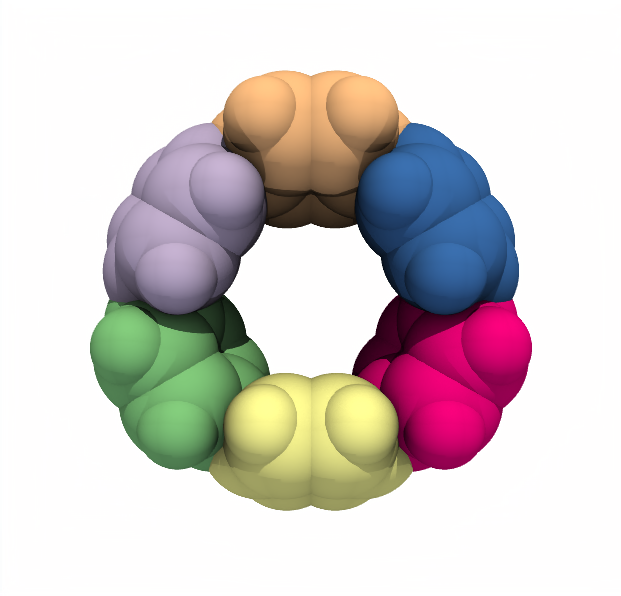}}
    \subfigure[Segmentation]{\includegraphics[width=0.32\columnwidth]{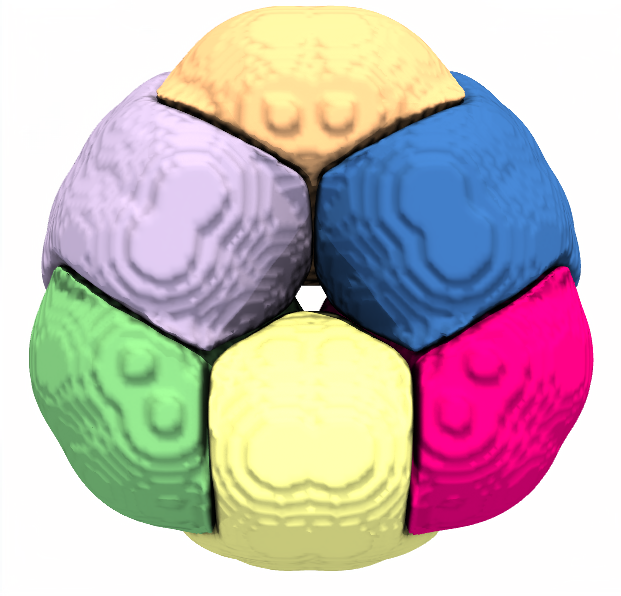}}
    \caption{[6]cycloparaphenylene. (a) van der Waals representation, (b) the six subgroups and (c) the Voronoi segmentation.}
    \label{fig:phe-cpp-atoms}
\end{figure}
\begin{figure*}[!t]
    \centering
    \begin{tabular}{c@{\hskip1pt}c@{\hskip1pt}c@{\hskip1pt}c@{\hskip1pt}c@{\hskip5pt}c}
    \toprule
    & \textbf{$\Phi_h$} & \textbf{$\Phi_p$} & \textbf{$q^d$} & \textbf{Transition diagram} & \textbf{Transition diagram}
    \\
    & {\tiny $-v$} \includegraphics[width=0.15\columnwidth]{Figures/isosurface_cmap.png} {\tiny $v\quad$}
    & {\tiny $-v$} \includegraphics[width=0.15\columnwidth]{Figures/isosurface_cmap.png} {\tiny $v\quad$}
    & {\tiny \quad Donor} \includegraphics[width=0.15\columnwidth]{Figures/red_blue_diverging.png} {\tiny Acceptor} 
    & (quadratic) 
    & (proportional)
    \\ 
    \midrule
    \raisebox{0.9\height}{\rotatebox{90}{\textbf{State 1}}}
    &
    %\subfigure[]{
    \includegraphics[width=0.35\columnwidth]{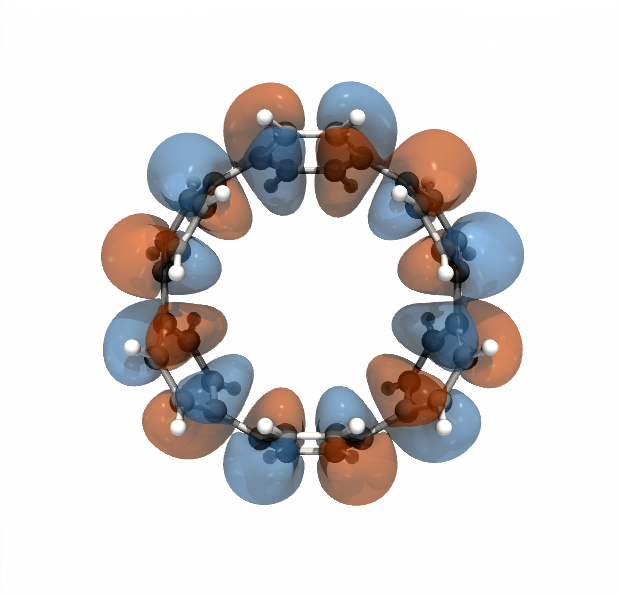}
    %}
    &
    %\subfigure[]{
    \includegraphics[width=0.35\columnwidth]{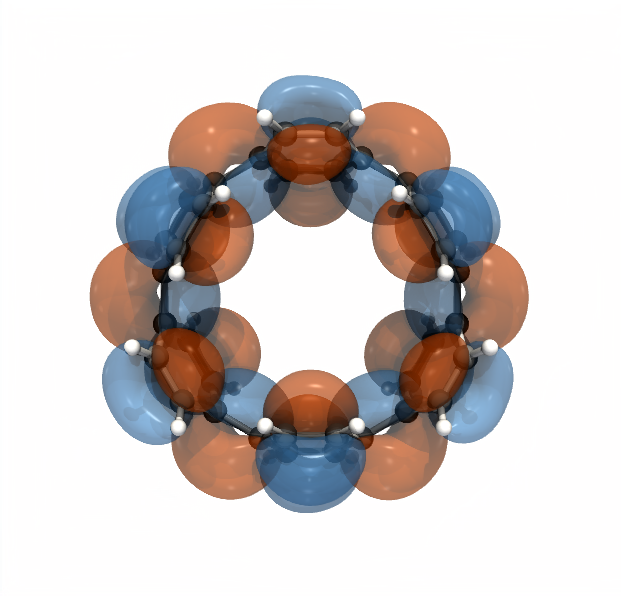}
    %}
    &
    %\subfigure[]{
    \includegraphics[width=0.35\columnwidth]{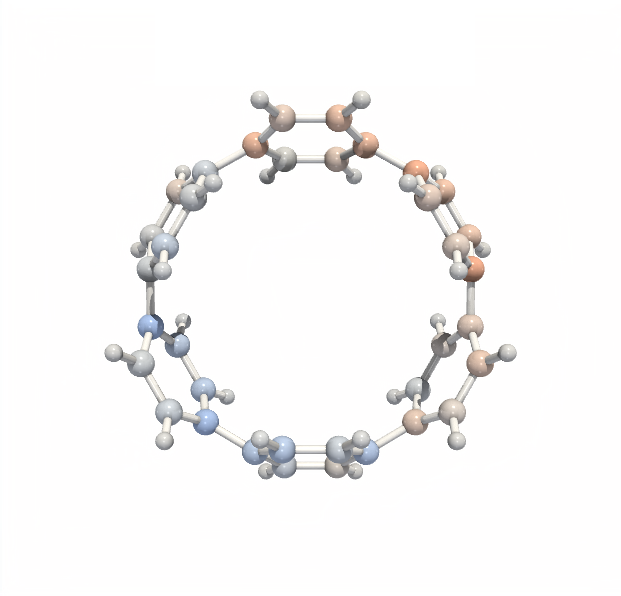}
    %}
    &
    %\subfigure[]{
    \includegraphics[height=0.35\columnwidth]{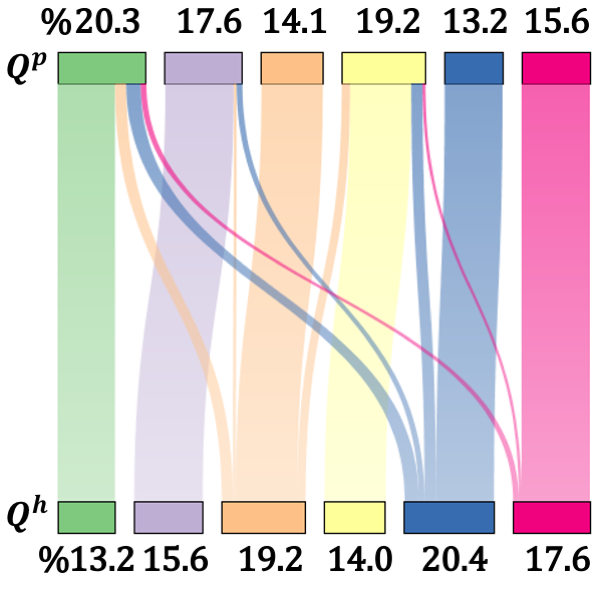}
    %}
    &
    %\subfigure[]{
    \includegraphics[height=0.35\columnwidth]{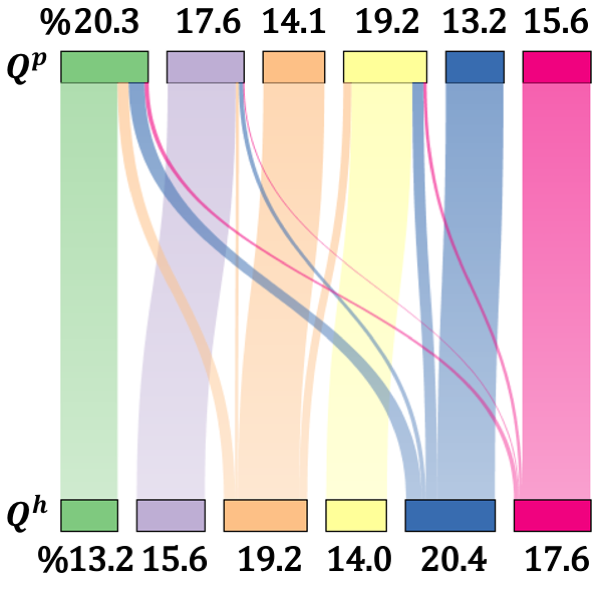}
    %}
    \\
    \raisebox{0.9\height}{\rotatebox{90}{\textbf{State 2}}}
    &
    %\subfigure[]{
    \includegraphics[width=0.35\columnwidth]{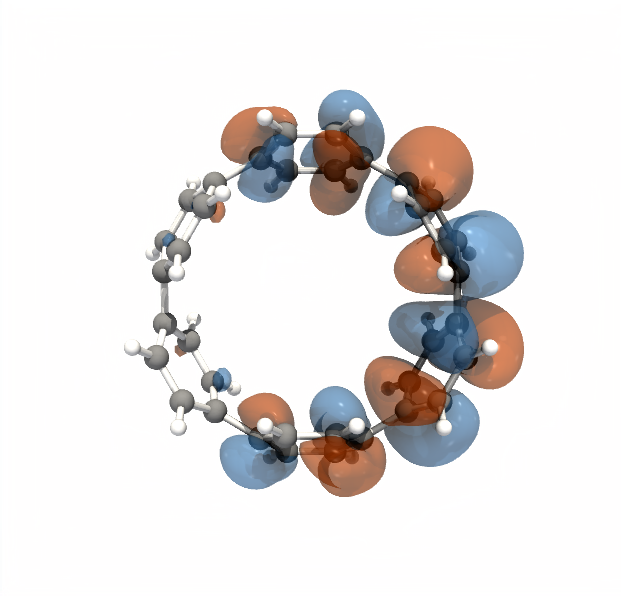}
    %}
    &
    %\subfigure[]{
    \includegraphics[width=0.35\columnwidth]{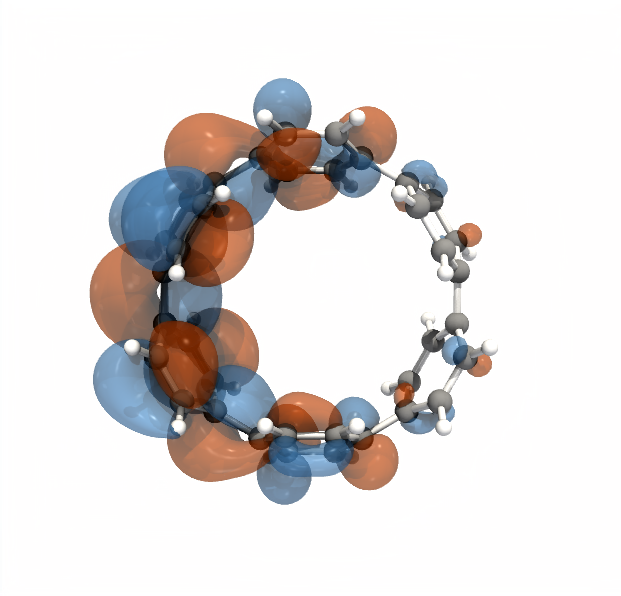}
    %}
    &
    %\subfigure[]{
    \includegraphics[width=0.35\columnwidth]{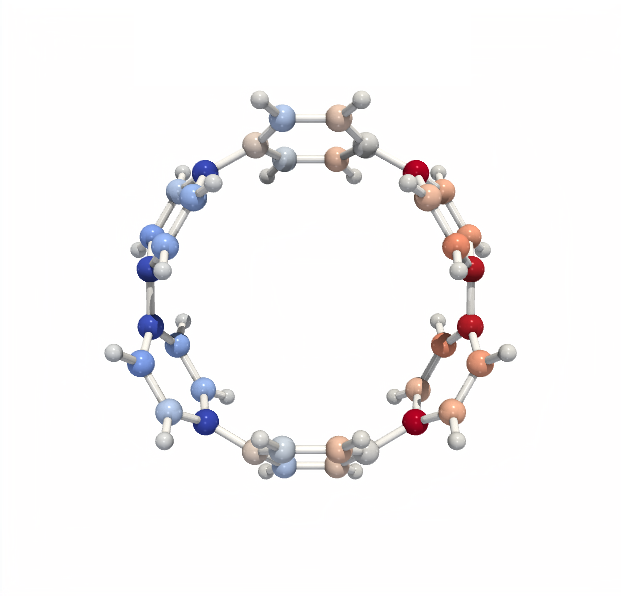}
    %}
    &
    %\subfigure[]{
    \includegraphics[height=0.35\columnwidth]{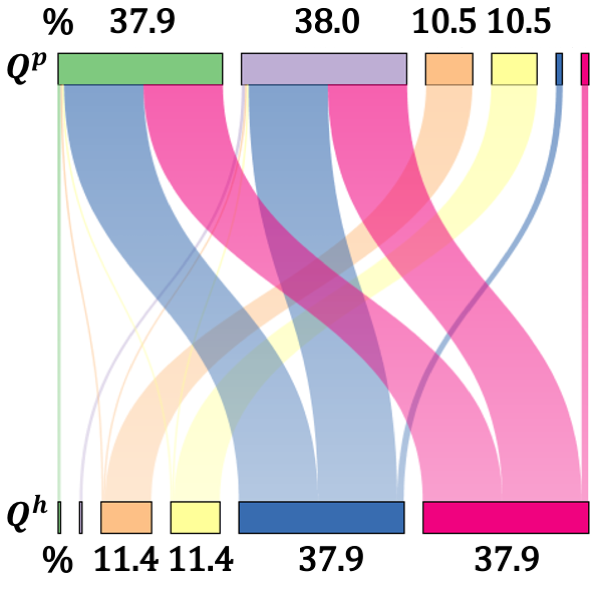}
    %}
    &
    %\subfigure[]{
    \includegraphics[height=0.35\columnwidth]{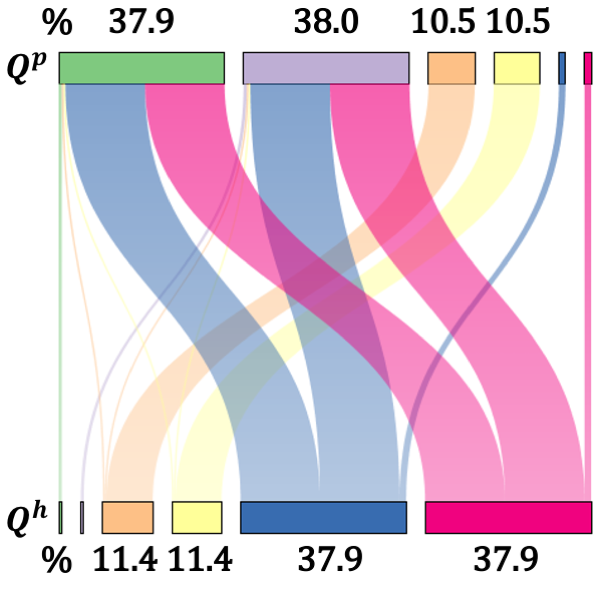}
    %}
    \\
    \raisebox{0.9\height}{\rotatebox{90}{\textbf{State 3}}}
    &
    %\subfigure[]{
    \includegraphics[width=0.35\columnwidth]{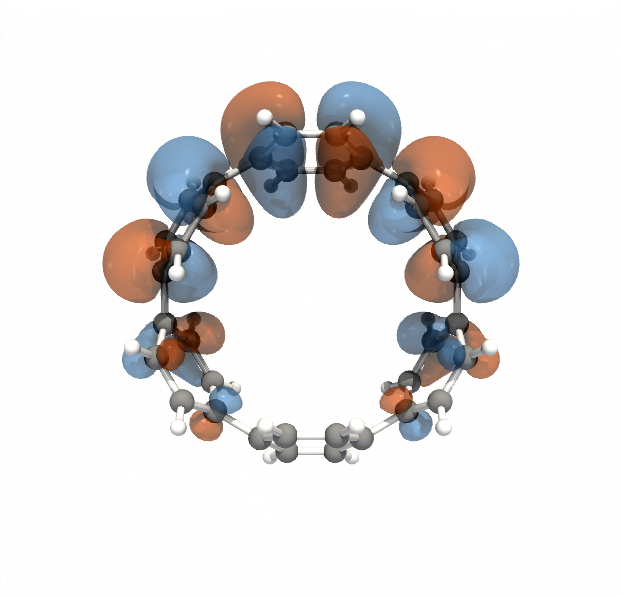}
    %}
    &
    %\subfigure[]{
    \includegraphics[width=0.35\columnwidth]{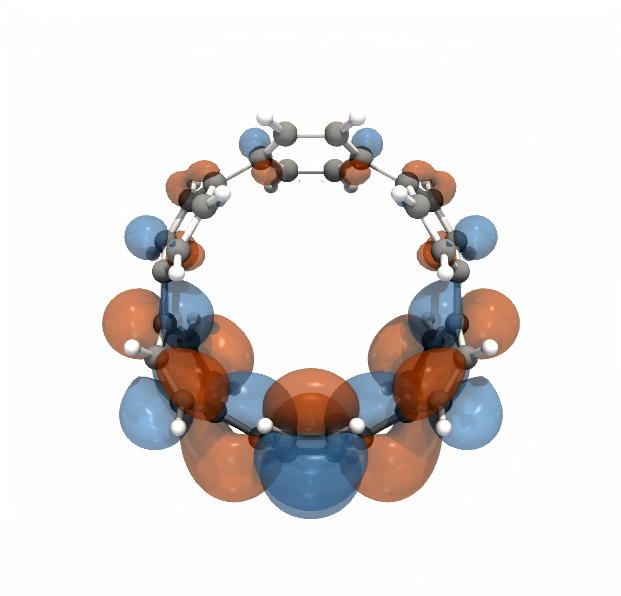}
    %}
    &
    %\subfigure[]{
    \includegraphics[width=0.35\columnwidth]{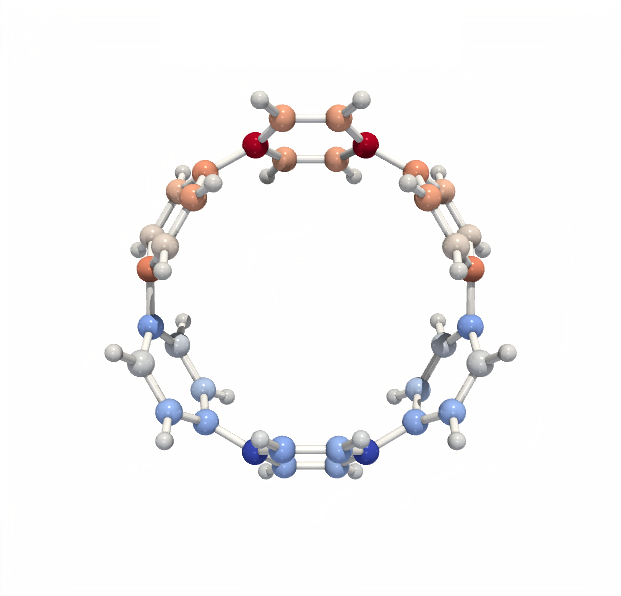}
    %}
    &
    %\subfigure[]{
    \includegraphics[height=0.35\columnwidth]{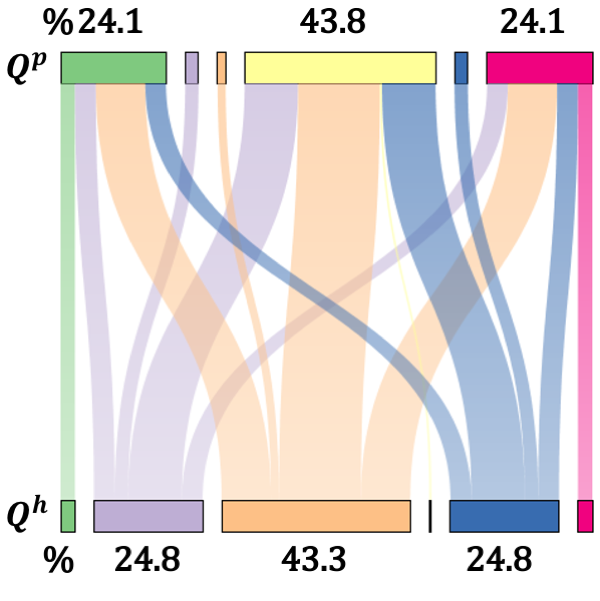}
    %}
    &
    %\subfigure[]{
    \includegraphics[height=0.35\columnwidth]{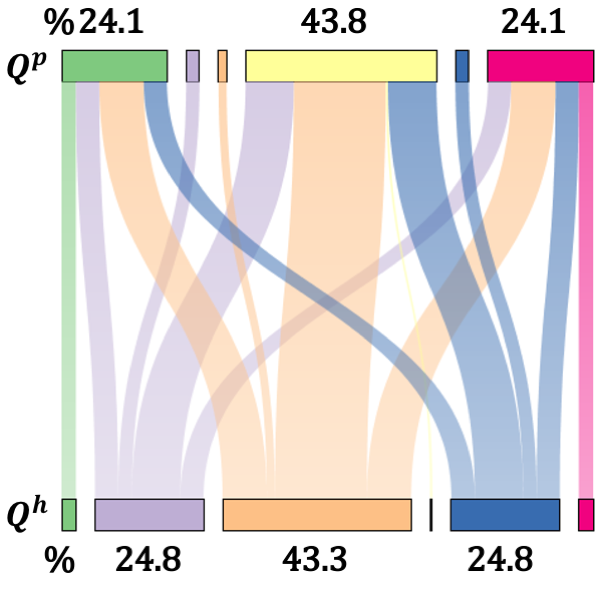}
    %}
    \end{tabular}
    \caption{[6]cycloparaphenylene. Cols. 1, 2 relate to task V1. They show selected isosurfaces for three different NTOs for the hole (col. 1) and particle (col. 2). Col. 3 shows results of task V2 displaying how the charge changes during the electronic excitation. Cols. 4 and 5 display results of task V3. They show results of the quadratic (col. 4) and proportional (col. 5) method for computing the transition diagram.}
    \label{fig:phe-cpp}
\end{figure*}

As the second case study, we investigated [6]cycloparaphenylene, a molecule composed of 6 benzene rings connected by covalent bonds in the para positions 
to form a ring Fig.~\ref{fig:phe-cpp-atoms}~\cite{cpp}. 
Note that for the other three case studies in this paper, the number of subgroups is at most three which implies the charge transfer problem has a unique solution. Here, the molecule is fragmented into six groups, the six phenyl rings, and thus provides a good study of a system having more than one donor and acceptor. In this case, the charge transfer problem is not uniquely determined, and the transition diagrams computed using proportional and quadratic approaches may differ. We are here focusing on the first three excited states~(Fig.~\ref{fig:phe-cpp}). While for the first state, both the hole and the particle are delocalized over the whole molecule; the second and third excited states present a clear charge transfer character. When looking at the NTO for those two states, they appear to be very similar. However, looking at the charge difference density and the transition diagrams, we can observe subtle differences such as the difference of localization of the particle NTO mainly shared on two cycles for the second excited state and three cycles for the third excited state. The difference observed between the proportional and quadratic transition diagrams are marginal for our case but might be important in other systems bearing a large number of subgroups.

%\begin{figure}[t]
%    \centering
%    \subfigure[Cu-PHE]{\includegraphics[width=0.32\columnwidth]{Figures/metal-phe/cu-phe2-State1_transitionDiagram.png}}
%    \subfigure[Ag-PHE]{\includegraphics[width=0.32\columnwidth]{Figures/metal-phe/ag-phe2-State1_transitionDiagram.png}}
%    \subfigure[Au-PHE]{\includegraphics[width=0.32\columnwidth]{Figures/metal-phe/au-phe2-State1_transitionDiagram.png}}
%    \caption{Transition diagrams for metal complexes. See Fig.~S1 in the supplement for a more detailed figure for this case study.}
%    \label{fig:metal-phe}
%\end{figure}

\begin{figure*}[t]
    \centering
    \begin{tabular}{c@{\hskip1pt}c@{\hskip1pt}c@{\hskip1pt}c@{\hskip1pt}c@{\hskip1pt}c}
    \toprule
    & \textbf{$\Phi_h$:} {\tiny $-v$} \includegraphics[width=0.12\columnwidth]{Figures/isosurface_cmap.png} {\tiny $v$} 
    & \textbf{$\Phi_p$:} {\tiny $-v$} \includegraphics[width=0.12\columnwidth]{Figures/isosurface_cmap.png} {\tiny $v$} 
    & \textbf{Segmentation} 
    & \textbf{$q^d$:} {\tiny Donor} \includegraphics[width=0.12\columnwidth]{Figures/red_blue_diverging.png} {\tiny Acceptor} 
    & \textbf{Transition diagram} 
    \\ 
    \midrule
    \raisebox{0.7\height}{\rotatebox{90}{\textbf{Cu-PHE}}}
    &
    %\subfigure[]{
    \includegraphics[width=0.35\columnwidth]{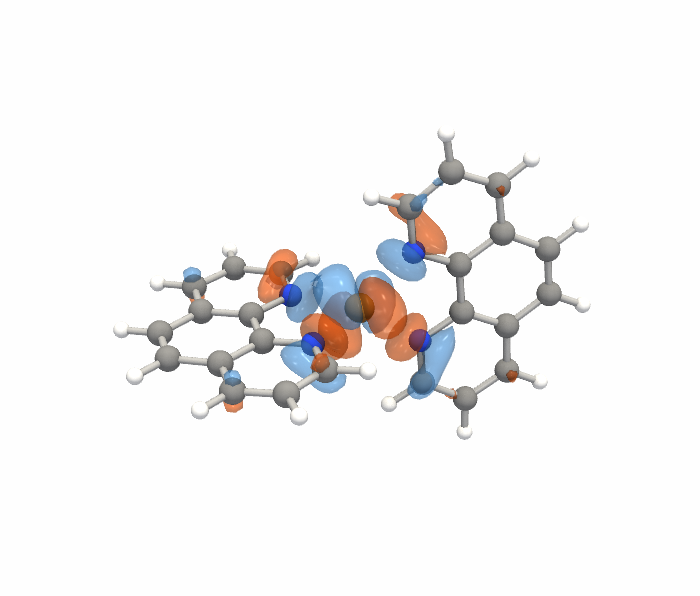}
    %}
    &
    %\subfigure[]{
    \includegraphics[width=0.35\columnwidth]{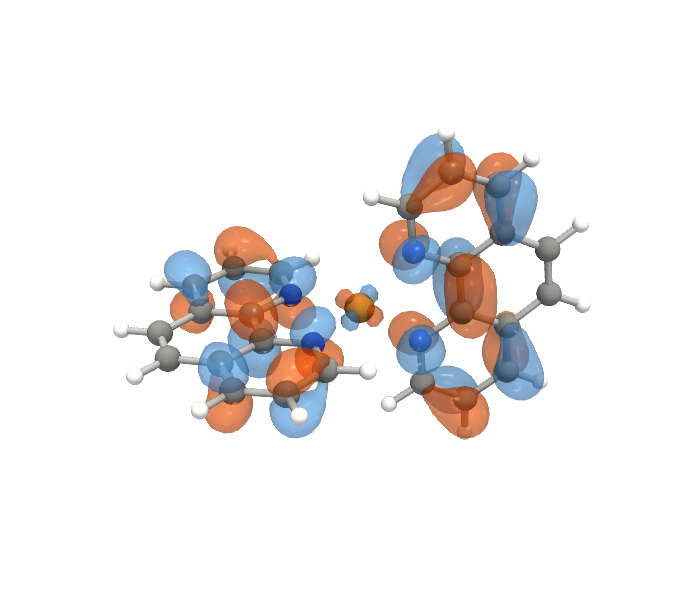}
    %}
    &
    %\subfigure[]{
    \includegraphics[width=0.35\columnwidth]{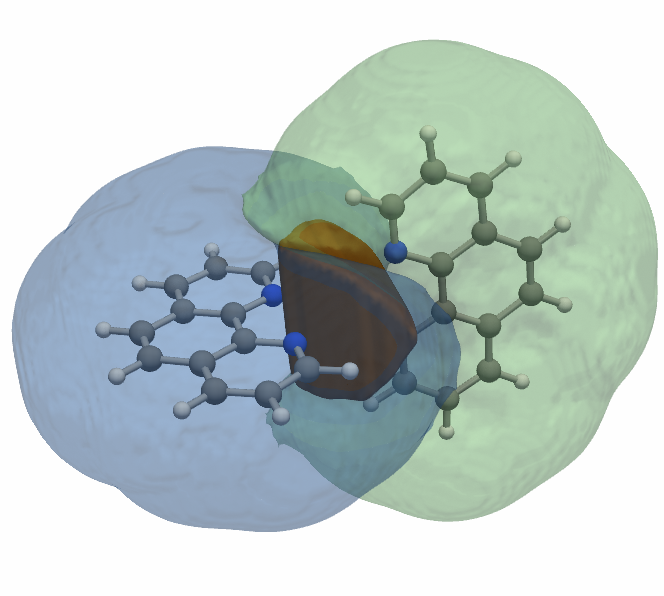}
    %}
    &
    %\subfigure[]{
    \includegraphics[width=0.35\columnwidth]{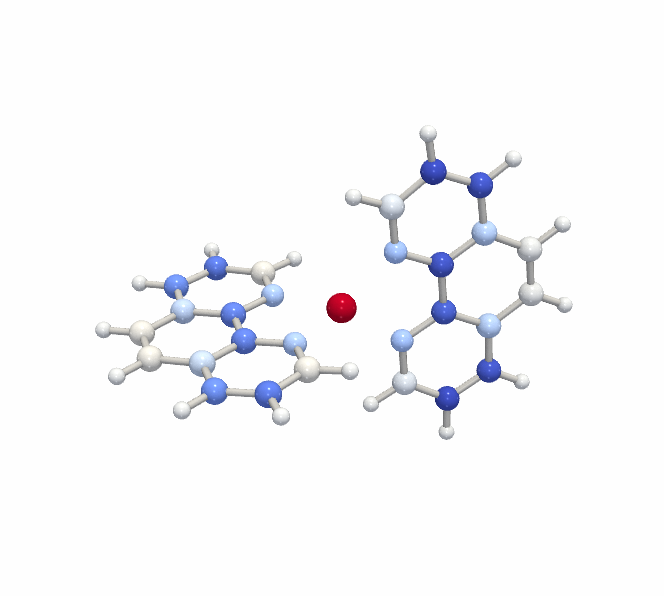}
    %}
    &
    %\subfigure[]{
    \includegraphics[height=0.35\columnwidth]{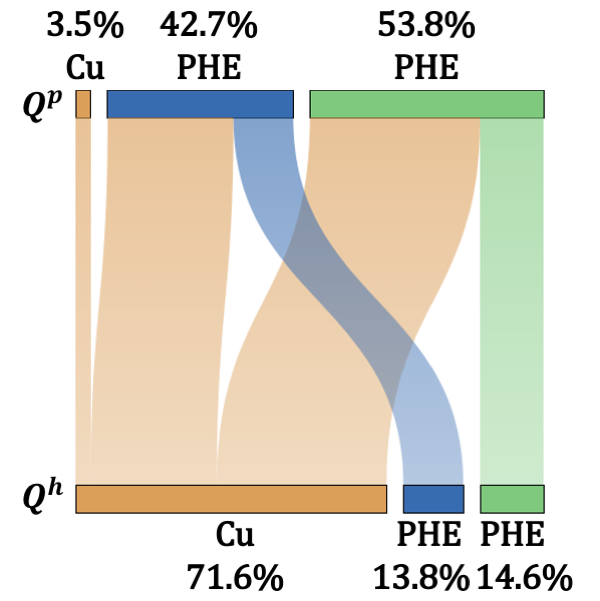}
    %}
    \\
    \raisebox{0.7\height}{\rotatebox{90}{\textbf{Ag-PHE}}}
    &
    %\subfigure[]{
    \includegraphics[width=0.35\columnwidth]{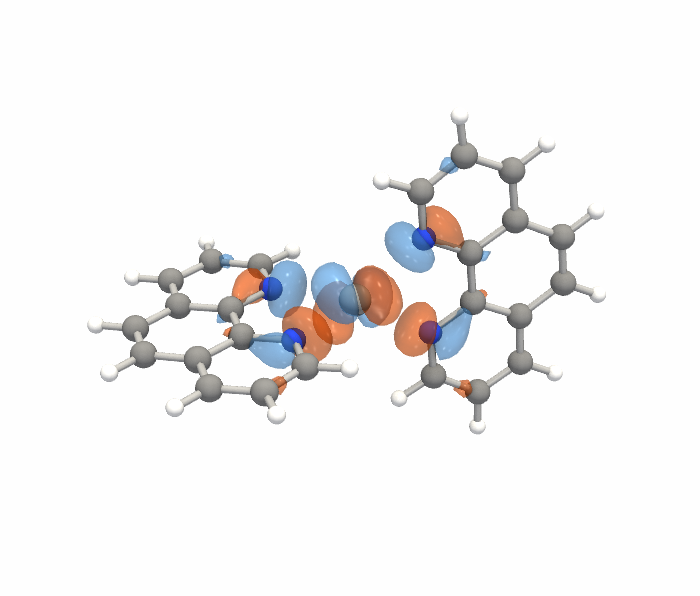}
    %}
    &
    %\subfigure[]{
    \includegraphics[width=0.35\columnwidth]{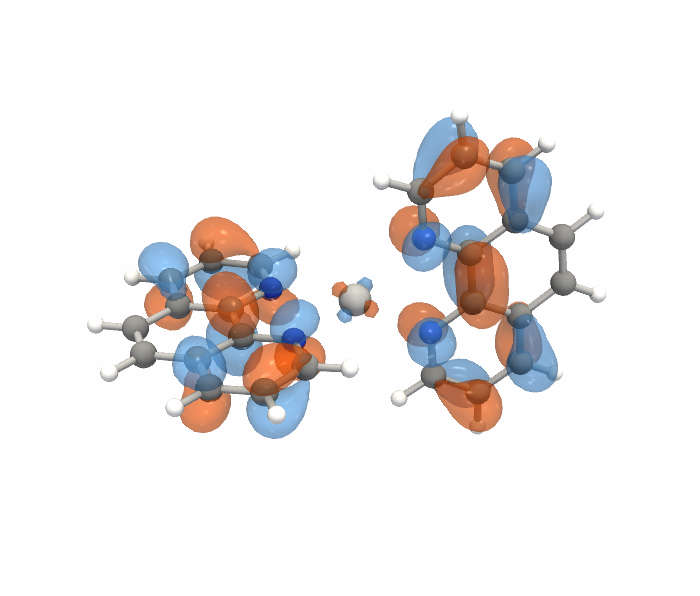}
    %}
    &
    %\subfigure[]{
    \includegraphics[width=0.35\columnwidth]{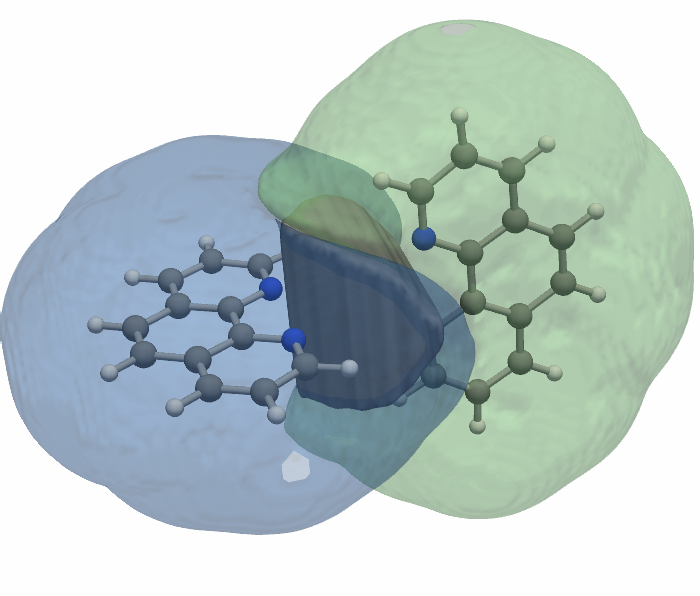}
    %}
    &
    %\subfigure[]{
    \includegraphics[width=0.35\columnwidth]{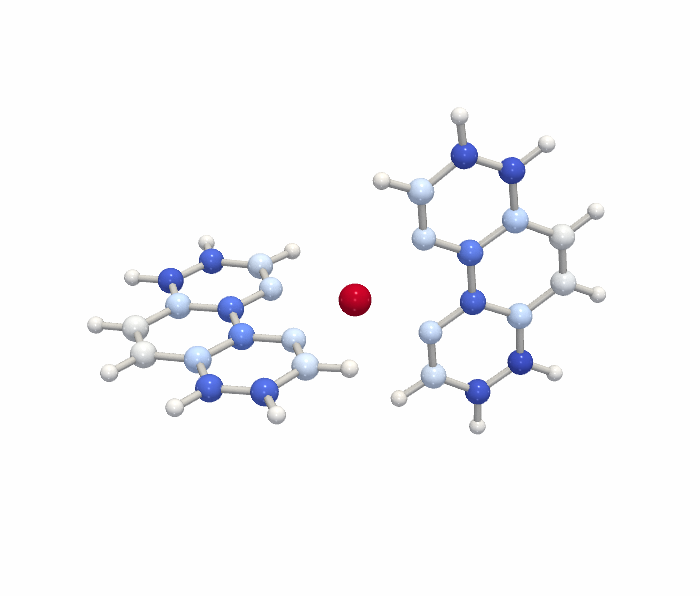}
    %}
    &
    %\subfigure[]{
    \includegraphics[height=0.35\columnwidth]{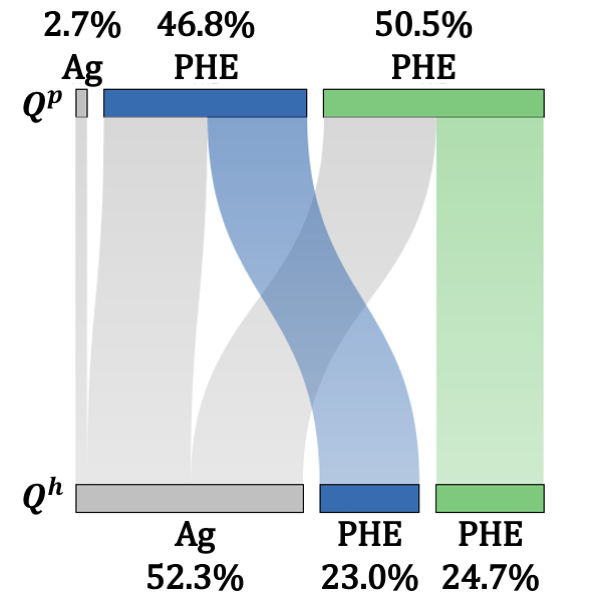}
    %}
    \\
    \raisebox{0.7\height}{\rotatebox{90}{\textbf{Au-PHE}}}
    &
    %\subfigure[]{
    \includegraphics[width=0.35\columnwidth]{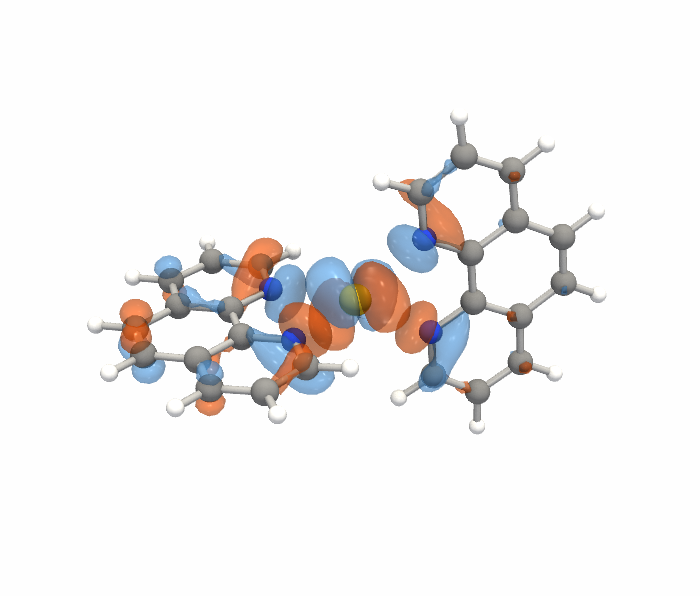}
    %}
    &
    %\subfigure[]{
    \includegraphics[width=0.35\columnwidth]{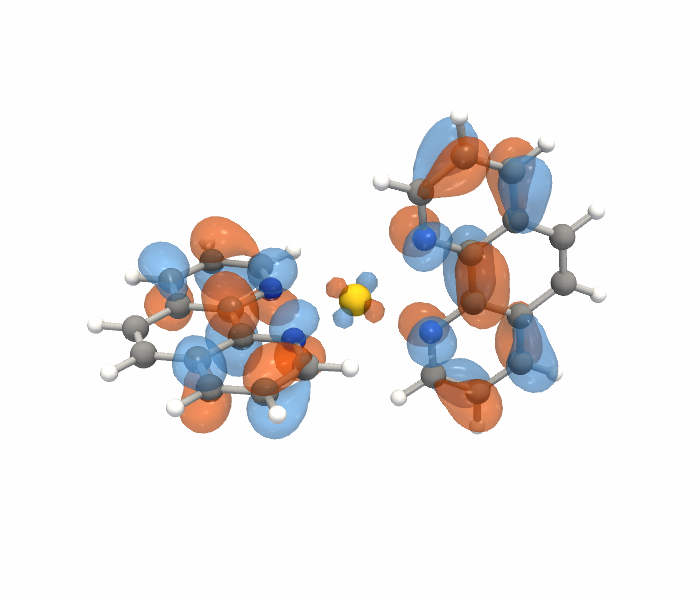}
    %}
    &
    %\subfigure[]{
    \includegraphics[width=0.35\columnwidth]{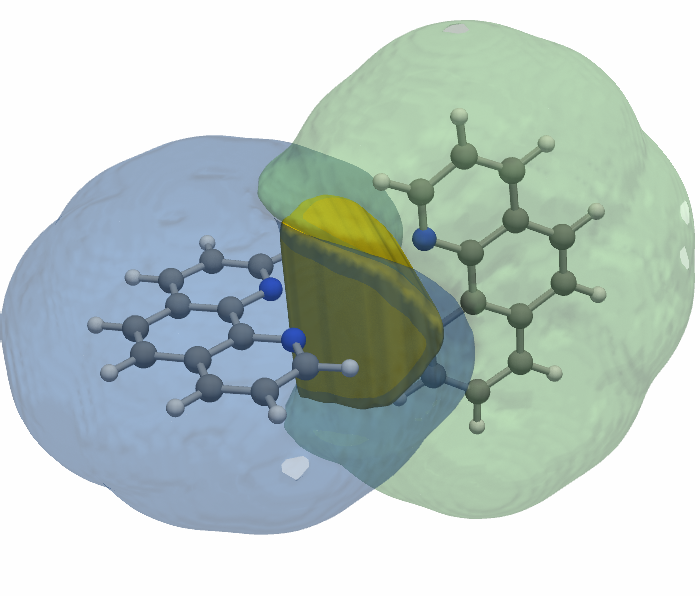}
    %}
    &
    %\subfigure[]{
    \includegraphics[width=0.35\columnwidth]{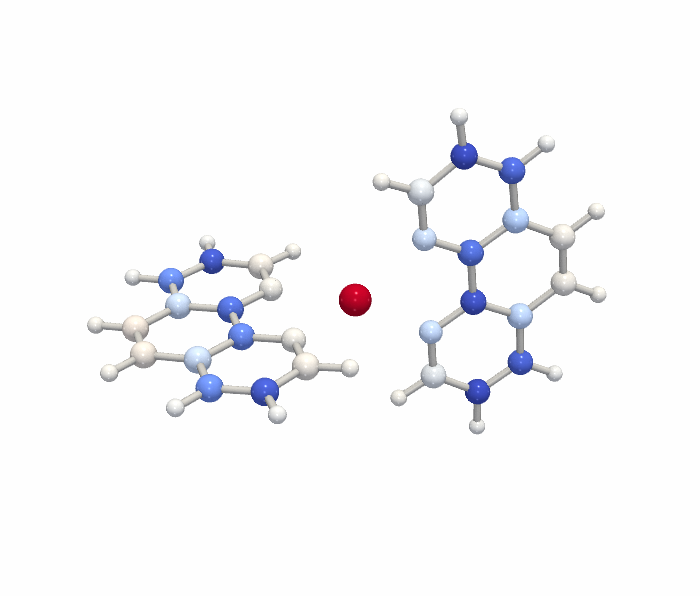}
    %}
    &
    %\subfigure[]{
    \includegraphics[height=0.35\columnwidth]{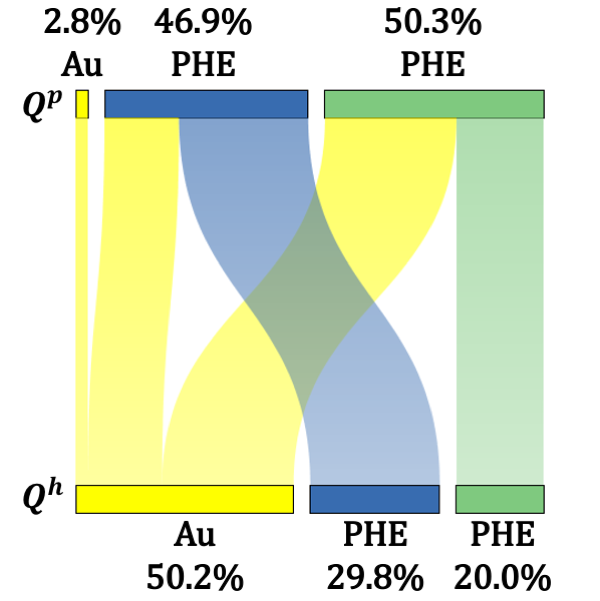}
    %}
    \end{tabular}
    \caption{Case study 3: Metal complexes. Cols. 1 and 2 relate to visualization task V1. They show the isosurfaces for the different NTOs for the hole (col. 1) and particle (col. 2). Col. 3 shows the segmentation of the different subgroups. Col. 4 shows the results of V2, displaying how the charge changes during the electronic excitation at atomic level of detail. Col. 5 displays the results of V3, showing the transition diagram for the different metal complexes.}
    \label{fig:metal-phe}
\end{figure*}

\subsection{Case study 3: Metal complexes}
Metal complexes present very interesting electronic structure properties and are used in a range of applications such as catalysis, organic solar cells, and organic light-emitting devices. They are composed of one or several metallic atoms and one or several molecules (called ligands) which organize around the metal atom. 
Quantum chemistry has been widely used to study these systems and to identify if specific electronic excitations are localized on a specific part of a molecule or if there is a transfer from the metal to a specific ligand. 
Identifying the nature of these excitations is a prerequisite for designing new molecules with specific properties. 
%As a case study 
We looked at the series of metal complexes made of one metal atom from group 11 (Copper \ce{Cu}, Silver \ce{Ag} and Gold \ce{Au}) and two phenanthroline (Phe) molecules arranged around it~(Fig.~\ref{fig:metal-phe}).
Here, we are focusing on the first excited states for the three metal complexes. 
As expected for complexes made with a metal atom from the same group, the particle charge distribution $Q^p$ for all three molecules is similar. 
However, the hole charge distribution $Q^h$ is different with \ce{Cu}-Phe having the most concentrated charge on \ce{Cu} ($71.6 \%$). In \ce{Ag}-Phe, $Q^h_{\ce{Ag}} = 52.3\%$ and in \ce{Au}-Phe, $Q^h_{\ce{Au}} = 50.2\%$. In \ce{Au}-Phe, the hole charge on the two Phe ligands is asymmetric ($29.8\%$ and $20.0\%$) as compared to the hole charge on Phe in \ce{Ag}-Phe ($23.0\%$ and $24.7\%$) and \ce{Cu}-Phe ($14.6\%$ and $13.8\%$). The transition diagrams allow to identify the differences between the three complexes with a larger charge transfer from the metal to the Phe for \ce{Cu}-Phe in comparison with \ce{Ag}-Phe and \ce{Au}-Phe.
    
\subsection{Case study 4: Copper complexes with various ligands} 
\begin{figure*}[t]
    \centering
    \begin{tabular}{c@{\hskip1pt}c@{\hskip1pt}c@{\hskip1pt}c@{\hskip1pt}c@{\hskip1pt}c}
    \toprule
    & \textbf{$\Phi_h$:} {\tiny $-v$} \includegraphics[width=0.12\columnwidth]{Figures/isosurface_cmap.png} {\tiny $v$}
    & \textbf{$\Phi_p$:} {\tiny $-v$} \includegraphics[width=0.12\columnwidth]{Figures/isosurface_cmap.png} {\tiny $v$}
    & \textbf{Segmentation} 
    & \textbf{$q^d$:} {\tiny Donor} \includegraphics[width=0.12\columnwidth]{Figures/red_blue_diverging.png} {\tiny Acceptor} 
    & \textbf{Transition diagram} 
    \\ 
    \midrule
    \raisebox{0.4\height}{\rotatebox{90}{\qquad\textbf{Phe}}}
    &
    %\subfigure[]{
    \includegraphics[width=0.35\columnwidth]{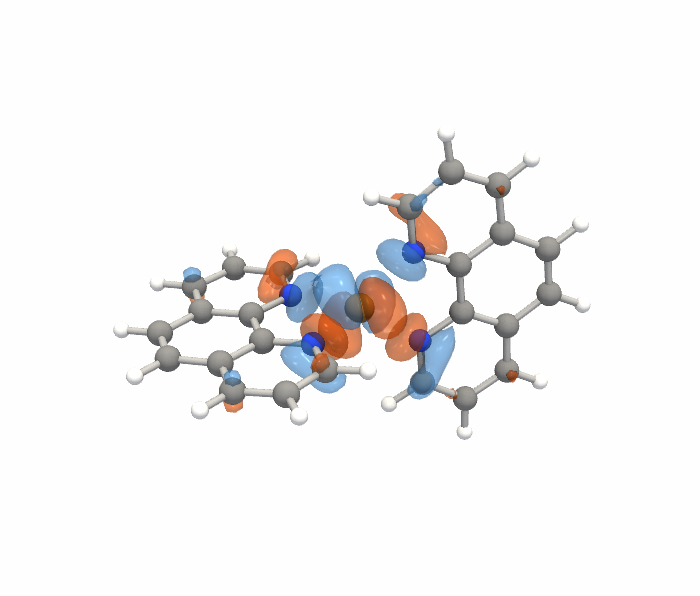}
    %}
    &
    %\subfigure[]{
    \includegraphics[width=0.35\columnwidth]{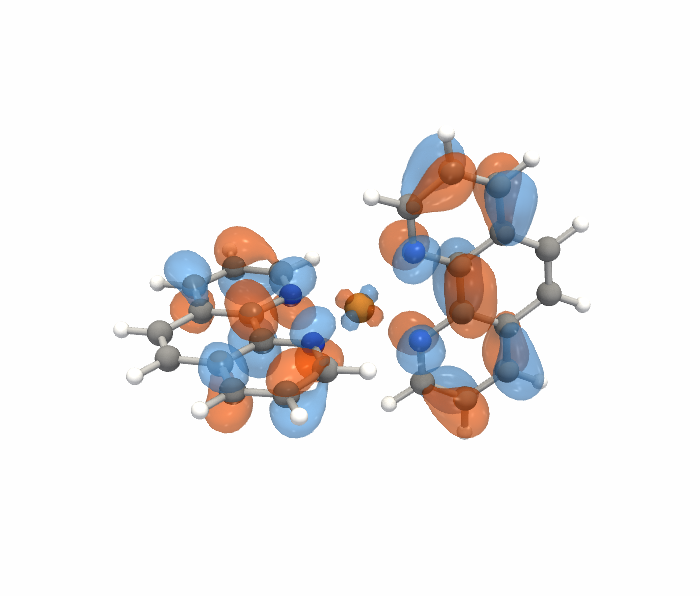}
    %}
    &
    %\subfigure[]{
    \includegraphics[width=0.35\columnwidth]{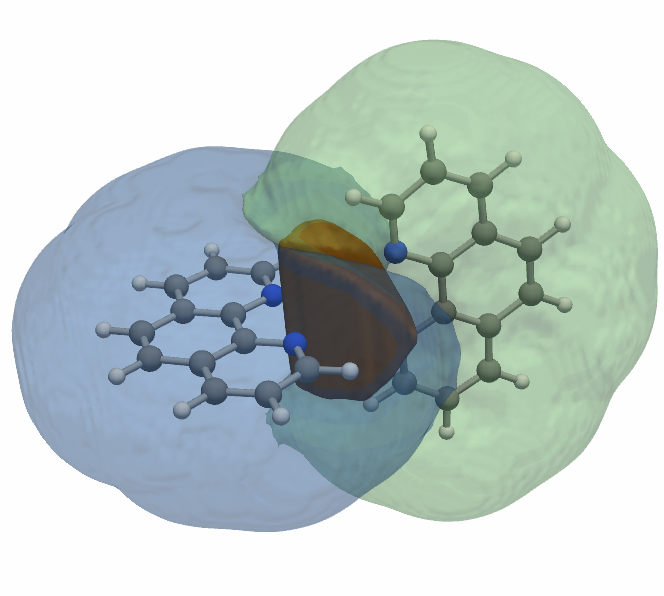}
    %}
    &
    %\subfigure[]{
    \includegraphics[width=0.35\columnwidth]{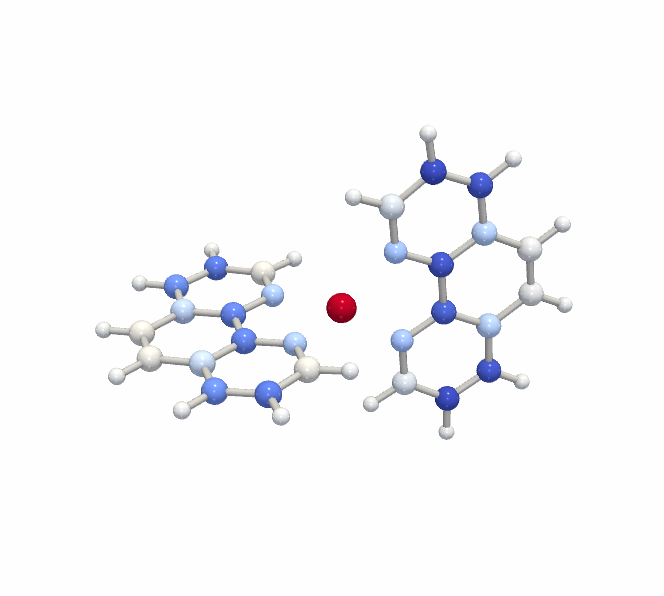}
    %}
    &
    %\subfigure[]{
    \includegraphics[width=0.35\columnwidth]{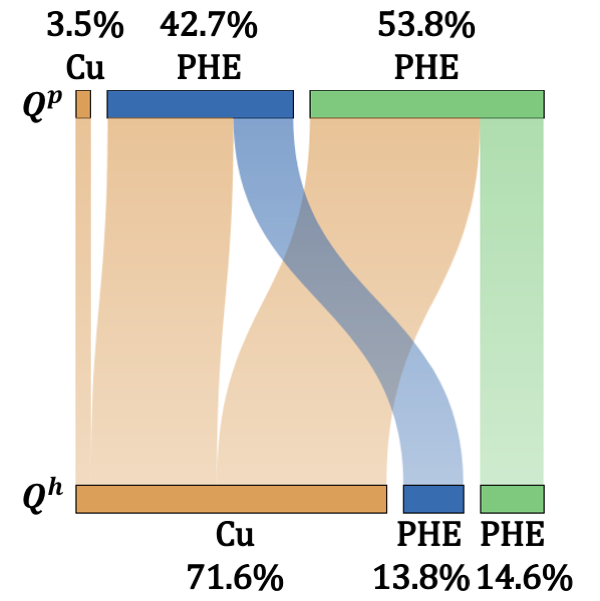}
    %}
    \\
    \raisebox{0.8\height}{\rotatebox{90}{\textbf{PhePhe}}}
    &
    %\subfigure[]{
    \includegraphics[width=0.35\columnwidth]{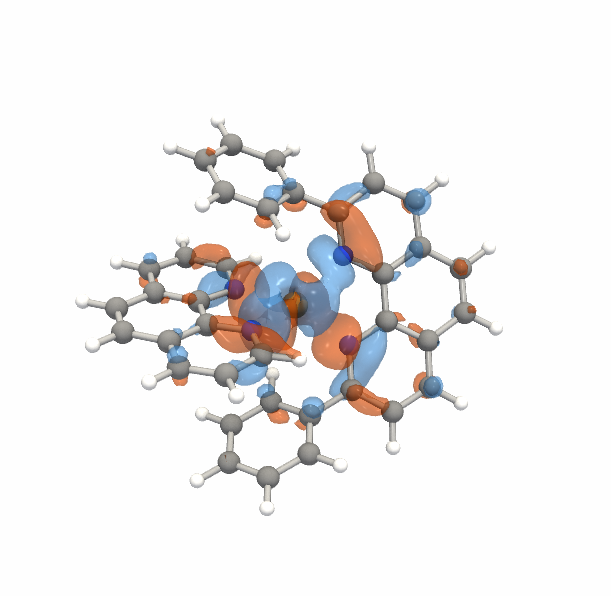}
    %}
    &
    %\subfigure[]{
    \includegraphics[width=0.35\columnwidth]{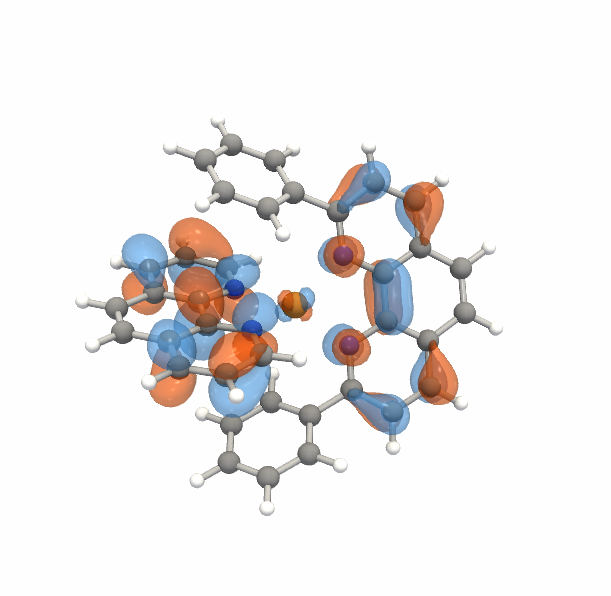}
    %}
    &
    %\subfigure[]{
    \includegraphics[width=0.35\columnwidth]{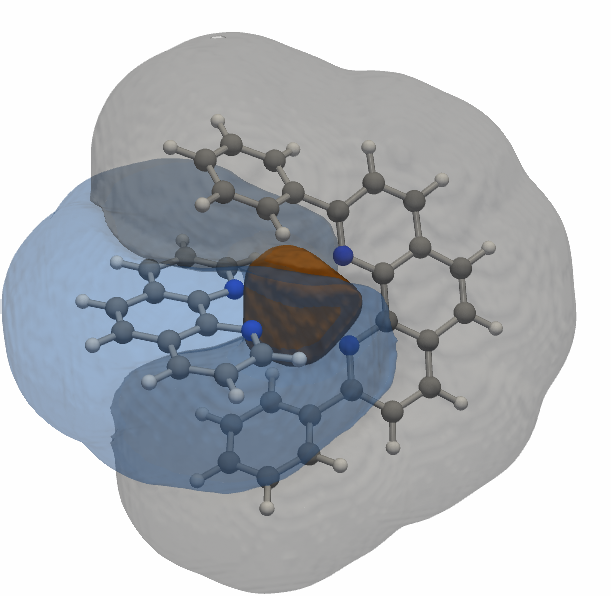}
    %}
    &
    %\subfigure[]{
    \includegraphics[width=0.35\columnwidth]{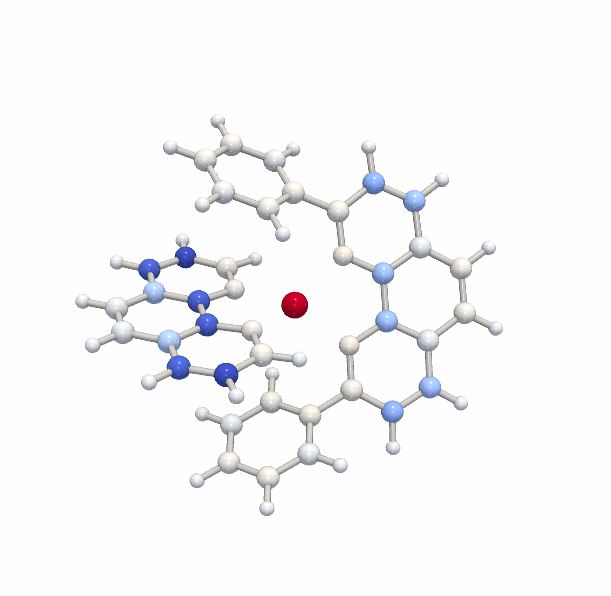}
    %}
    &
    %\subfigure[]{
    \includegraphics[width=0.35\columnwidth]{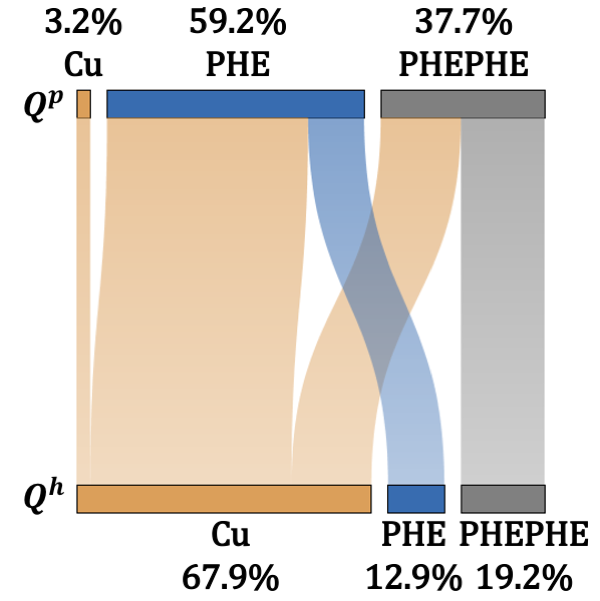}
    %}
    \\
    %\midrule
   \raisebox{0.8\height}{\rotatebox{90}{\textbf{Pheme}}}
    &
    %\subfigure[]{
    \includegraphics[width=0.35\columnwidth]{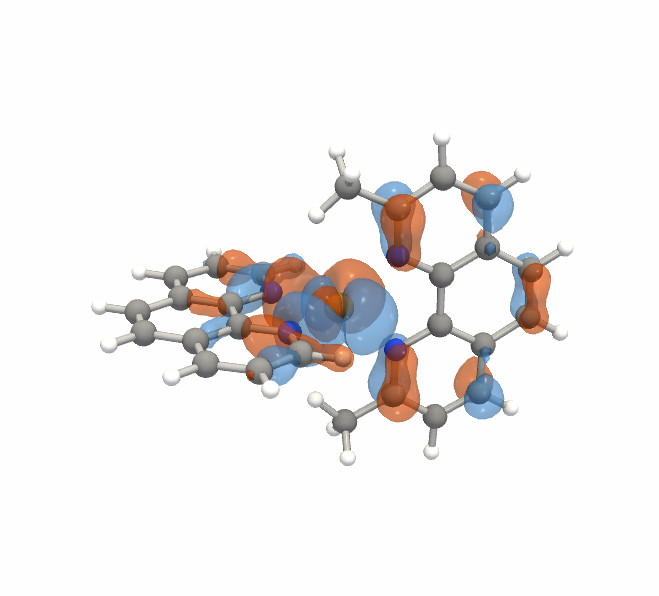}
    %}
    &
    %\subfigure[]{
    \includegraphics[width=0.35\columnwidth]{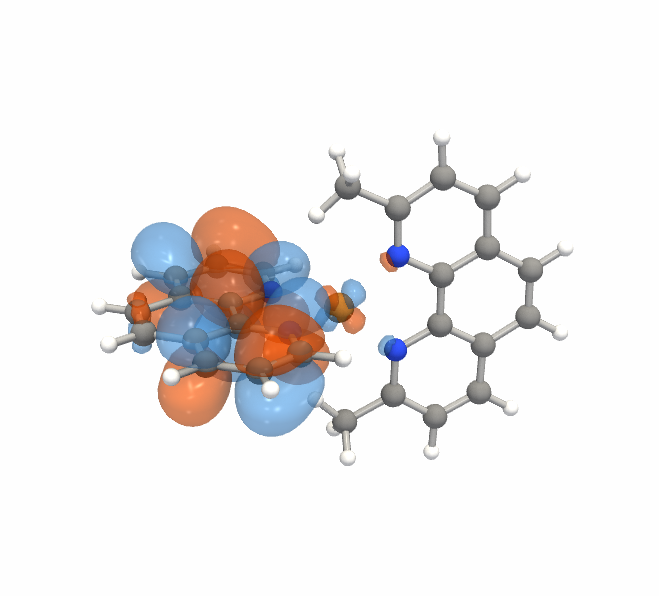}
    %}
    &
    %\subfigure[]{
    \includegraphics[width=0.35\columnwidth]{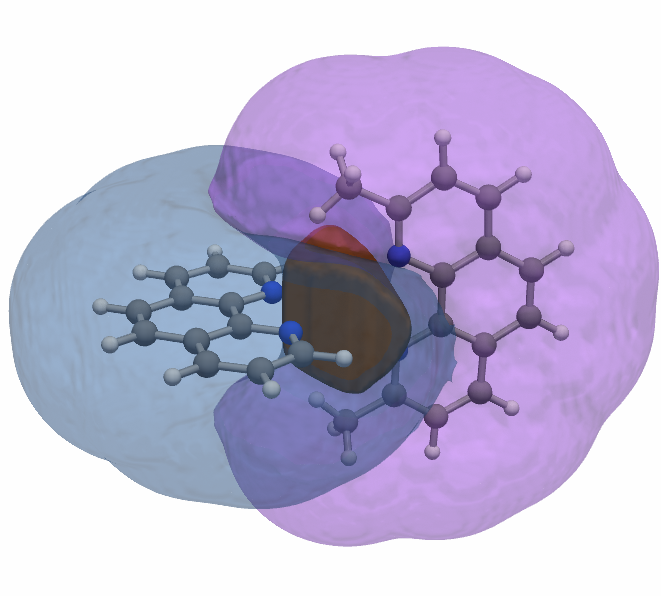}
    %}
    &
    %\subfigure[]{
    \includegraphics[width=0.35\columnwidth]{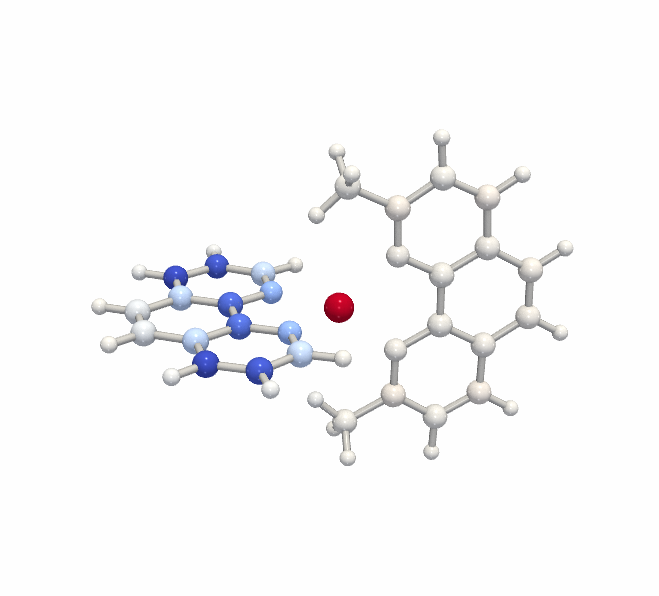}
    %}
    &
    %\subfigure[]{
    \includegraphics[width=0.35\columnwidth]{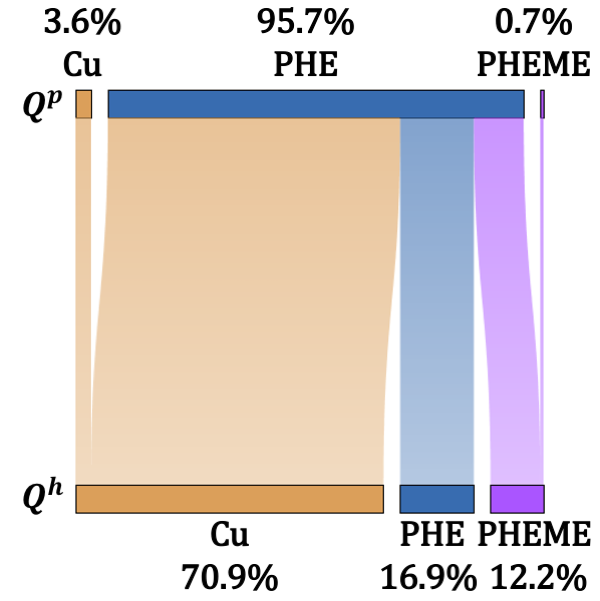}
    %}
    \\
    \raisebox{0.65\height}{\rotatebox{90}{\textbf{Pheome}}}
    &
    %\subfigure[]{
    \includegraphics[width=0.35\columnwidth]{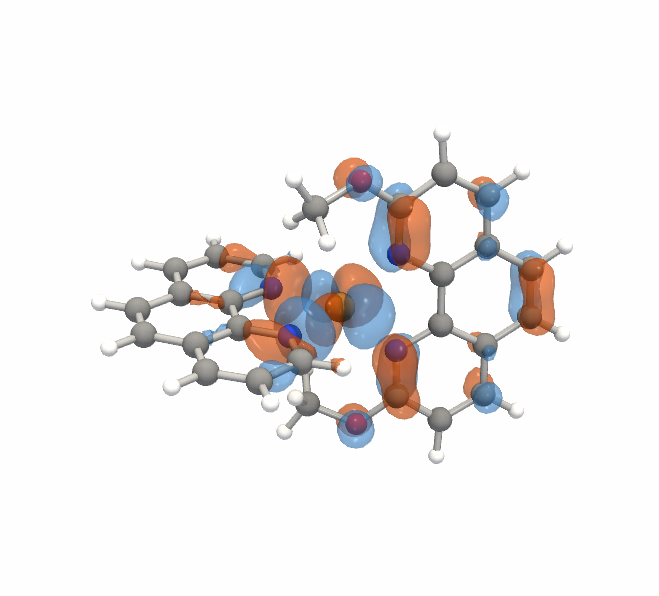}
    %}
    &
    %\subfigure[]{
    \includegraphics[width=0.35\columnwidth]{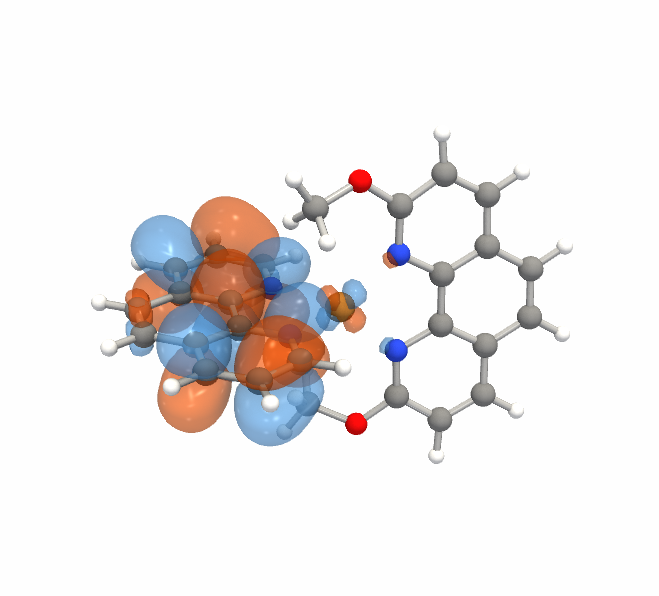}
    %}
    &
    %\subfigure[]{
    \includegraphics[width=0.35\columnwidth]{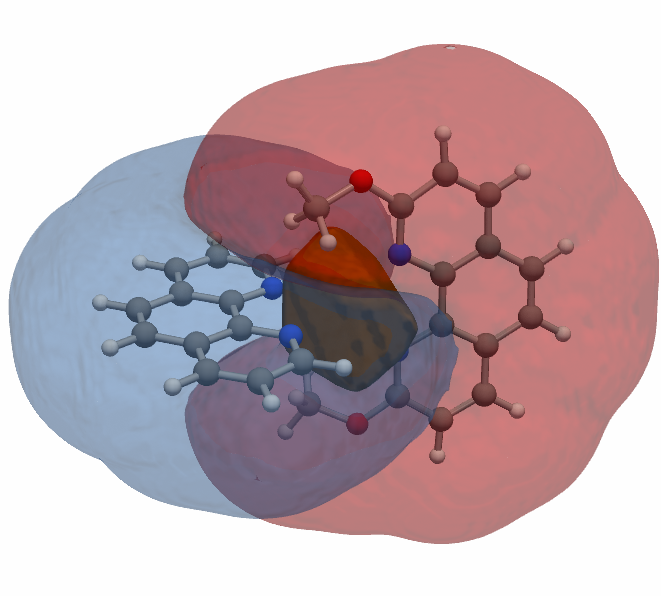}
    %}
    &
    %\subfigure[]{
    \includegraphics[width=0.35\columnwidth]{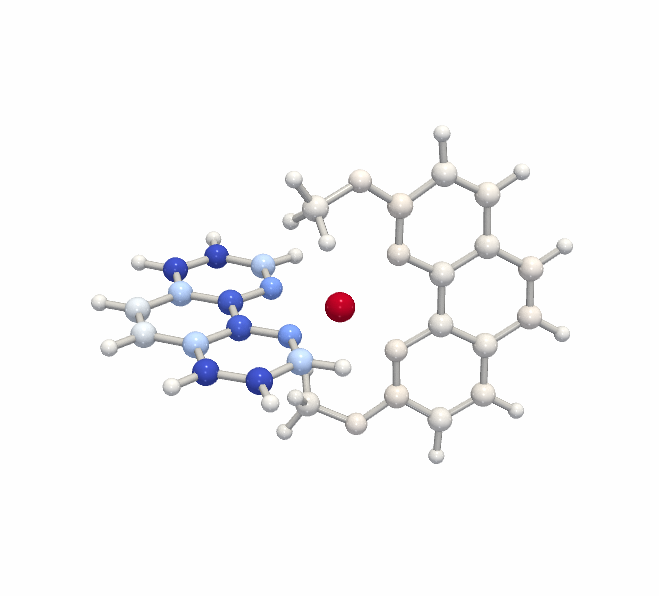}
    %}
    &
    %\subfigure[]{
    \includegraphics[width=0.35\columnwidth]{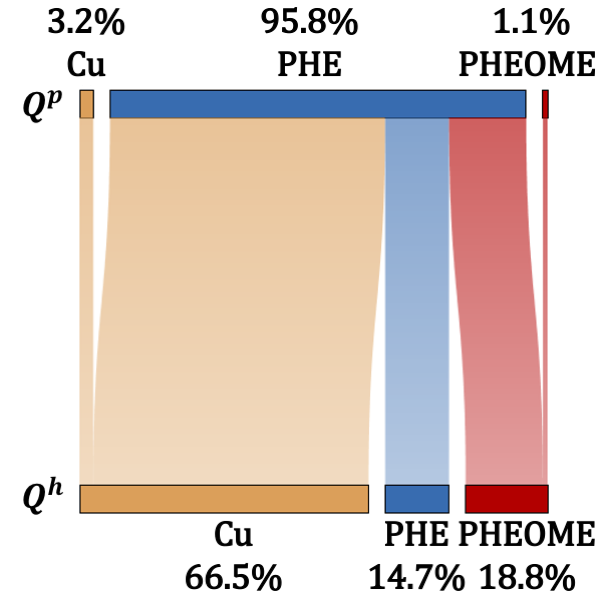}
    %}
    \\
    \raisebox{0.65\height}{\rotatebox{90}{\quad\textbf{IPR}}}
    &
    %\subfigure[]{
    \includegraphics[width=0.35\columnwidth]{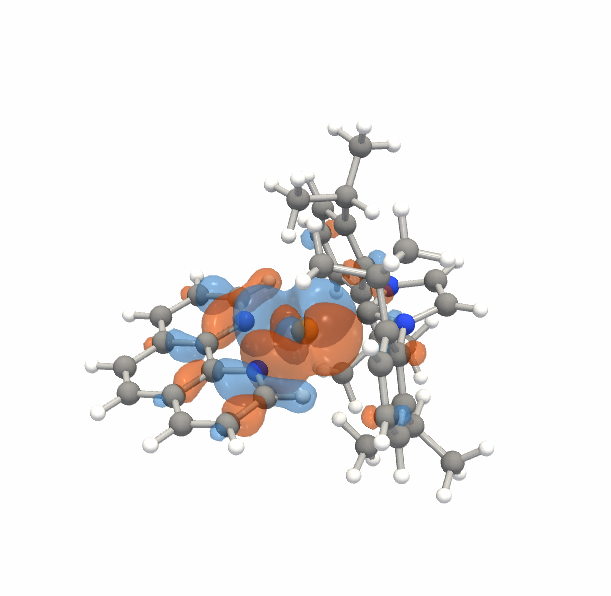}
    %}
    &
    %\subfigure[]{
    \includegraphics[width=0.35\columnwidth]{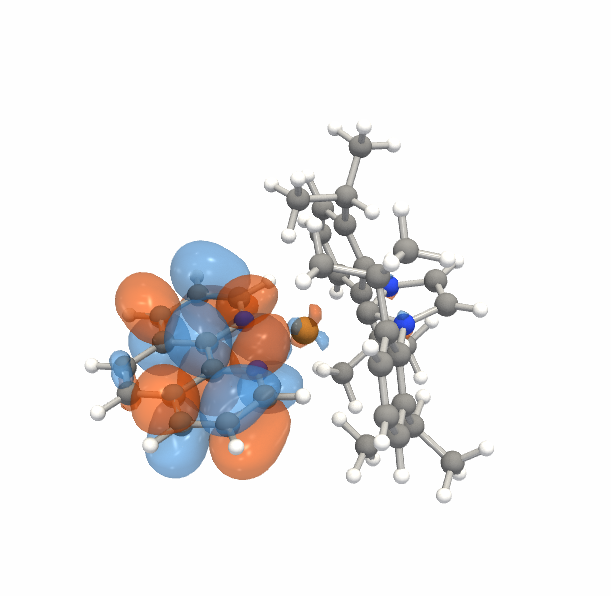}
    %}
    &
    %\subfigure[]{
    \includegraphics[width=0.35\columnwidth]{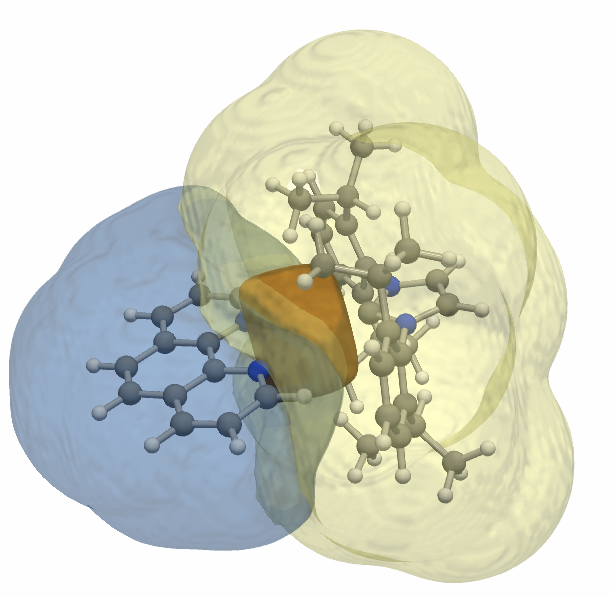}
    %}
    &
    %\subfigure[]{
    \includegraphics[width=0.35\columnwidth]{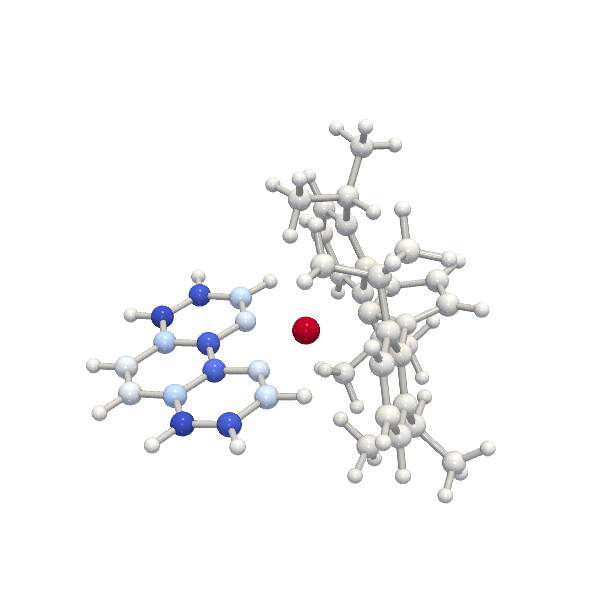}
    %}
    &
    %\subfigure[]{
    \includegraphics[width=0.35\columnwidth]{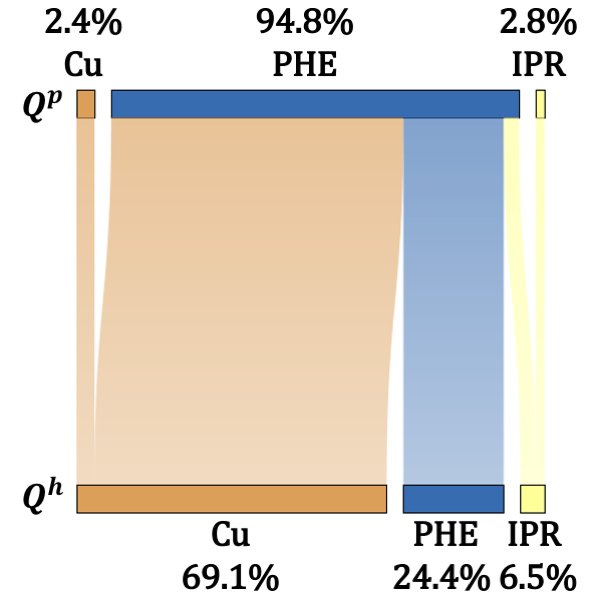}
    %}
    \\
    \raisebox{0.6\height}{\rotatebox{90}{\quad\textbf{XANT}}}
    &
    %\subfigure[]{
    \includegraphics[width=0.35\columnwidth]{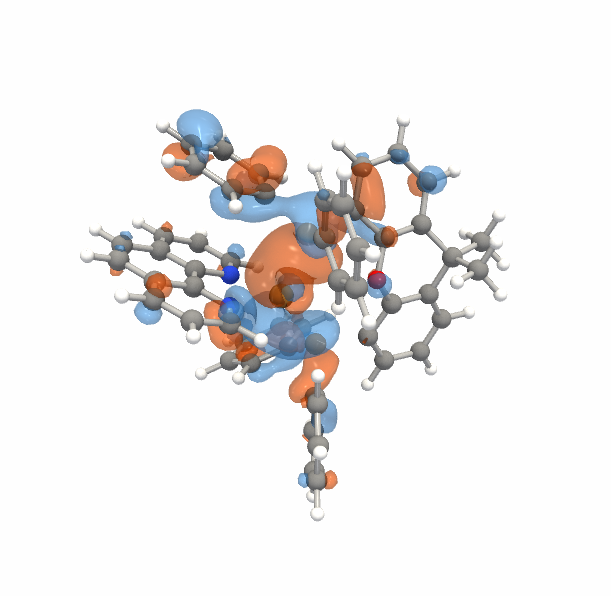}
    %}
    &
    %\subfigure[]{
    \includegraphics[width=0.35\columnwidth]{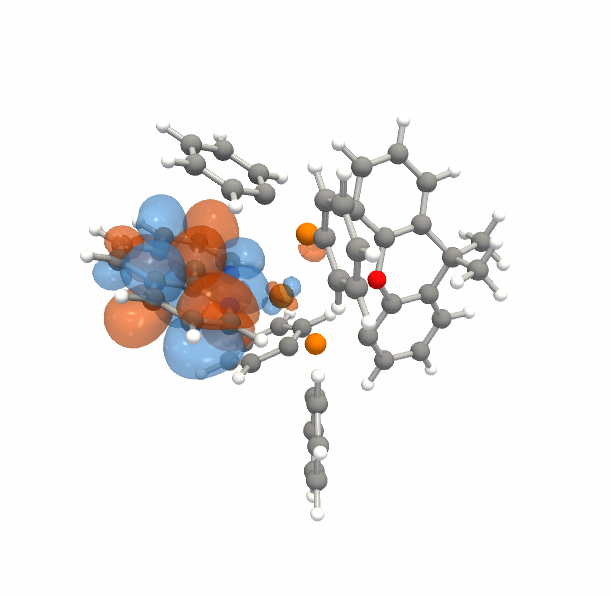}
    %}
    &
    %\subfigure[]{
    \includegraphics[width=0.35\columnwidth]{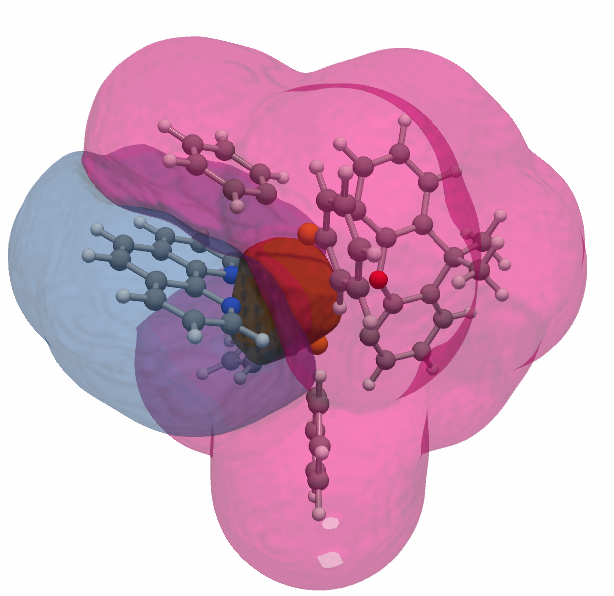}
    %}
    &
    %\subfigure[]{
    \includegraphics[width=0.35\columnwidth]{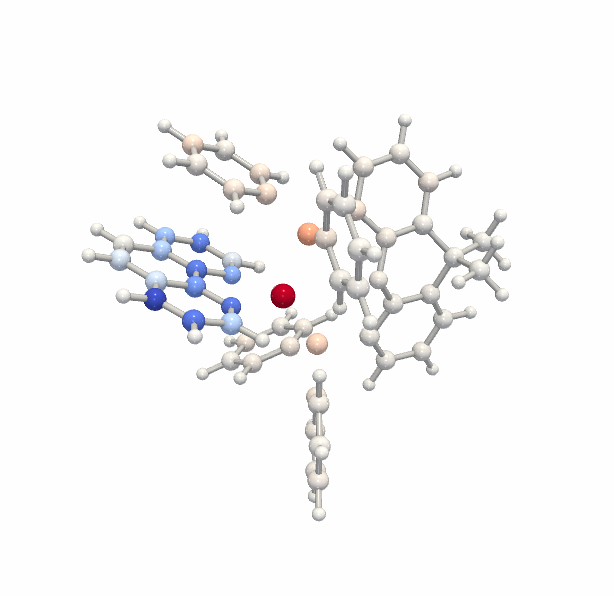}
    %}
    &
    %\subfigure[]{
    \includegraphics[width=0.35\columnwidth]{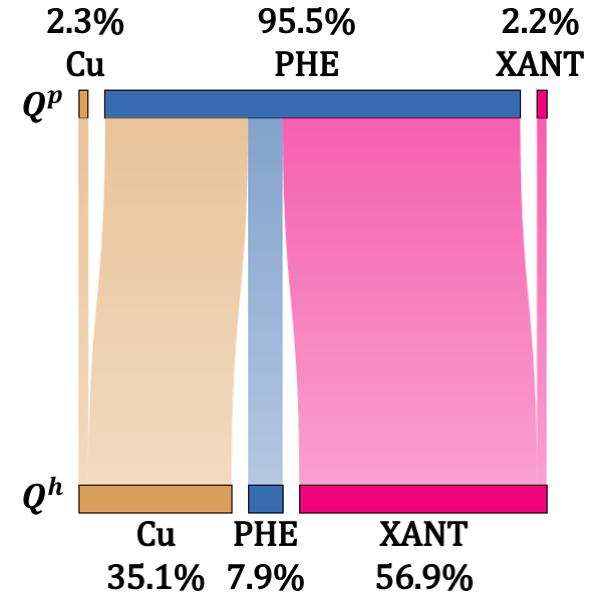}
    %}
    \end{tabular}
    \caption{Case study 4: Copper complexes with various ligands. Columns 1 and 2 show selected isosurfaces for the hole and particle NTOs respectively. Column 3 shows the volume segmentation for the different subgroups. Column 4 shows how the charge changes during the electronic excitation. The last column shows the transition diagram for the different ligands. Notice how the transition diagrams in top two rows~(\emph{Phe} and \emph{PhePhe}) are similar and how they differ from the second category, the next four rows.}
    \label{fig:cu-ligand}
\end{figure*}

We investigate a series of copper complexes bearing two ligands. The first ligand phenanthroline~(Phe) is always the same in all complexes, while the second ligand varies: 2,9-diphenyl-phenanthroline (PhePhe), 2,9-dimethyl-phenanthroline (Pheme), 2,9-dimethoxy-phenanthroline (Pheome), a N-heterocyclic Carbene~(IPR), and a diphosphine ligand~(XANT). 
We are focusing here on the first excited states. 
The transition diagrams presented in Fig.~\ref{fig:cu-ligand} enable identification of the differences between the copper complexes as for Cu-Phe and Cu-Phe-PhePhe, the charge transfer occurs from the copper atom to both ligands. However, for the four other complexes, the charge transfer to Phe is coming from both the copper atom and the other ligands. Moreover, we observe that Cu-Phe-XANT presents a unique behavior since more of the transfer comes from the diphosphine ligand than from the Copper atom. This illustrates how this approach can be used on a series of molecules to identify quickly the typical differences between them. 

\section{Discussion}
\label{sec:discussion}
Here we discuss some of the visualization design decisions and report the domain expert feedback. Discussion on the robustness and run-time performance of the method follows afterwards. 
\paragraph*{Visualization design.}
Our framework has been developed with computational chemists in the loop with constant feedback. During these discussions and along with the first experiments, the visualization and analysis tasks listed in Sec.~\ref{sec:problem} were developed. 
The process started with designs using bar charts which did not show the charge transfer. On the way to the final visualization design, we experimented with, among other visual representations, chord diagrams which were difficult to interpret for the chemists. Finally, we settled on Sankey diagram which is a simple yet effective visualization intuitively fulfilling all our visualization requirements. The transition diagram displays the charge distributions of hole and particle in the bottom and top bars and simplifies the comparison of the distributions. Showing the hole charge distribution at the bottom and particle at the top was natural for chemists because of their familiarity with energy diagrams, as against the standard practice of showing the flow in a Sankey diagram from left to right. Further, the transition diagram clearly distinguishes localized and non-localized charge transfer, a key requirement for chemists. In retrospect, we can summarize the design guidelines as providing a simple and direct visualization restricted to the most relevant aspects.

\paragraph*{User feedback.}
In summary, the user feedback on the final result was very positive. There was an agreement that this visualization fills a gap in the analysis tools typically used for this task.
The visualization gives easy access to quantitative information about the electronic transition that can be hardly derived from the side-by-side comparison of isosurfaces, a method that is typically used in such scenarios.
The four use cases discussed above represent real-world examples, both in their type and complexity.
While appreciating the presented methods, the computational chemists also stimulated discussion of further developments with respect to an increasing number and complexity of the configurations investigated, which we briefly discuss next in Section~\ref{sec:conclusion}.

\paragraph*{Robustness.} 
As evident from the presented case studies and user feedback, the transition diagrams present the complete information about the subgroup charges and charge transfer very well for a small number of subgroups. They also capture the similarities and differences in transitions for a given set of molecules, even when the order of subgroups is changed as long as the order is changed consistently for all the molecules. However, further investigation is needed to study the scalability of these diagrams with an increasing number of subgroups. Novel visualization techniques may be needed to address these challenges.

One of the advantages of using the weighted Voronoi diagram-based segmentation method is that it is dependent only on the atomic positions and radii, and not on the density field or its quality. However, the grid resolution does affect the segmentation quality in a discrete setting. The effect of the grid resolution on the quality of segmentation and computed charges need further investigation, but it is easy to see that higher resolution grids would be required for molecules containing a larger number of atoms. We also compared the weighted Voronoi-based segmentation with a gradient-based segmentation and found the computed charges at the subgroup level to be very similar, see the appendix~(Table~\ref{table:comparison}) for details.
  
\paragraph*{Run time analysis.} 
We performed some experiments to study the run time performance of our implementation and found that the segmentation was the most time-consuming stage of the pipeline. Although efficient implementation was not the focus of this work, we exploited parallelism during the segmentation and charge computation stages of the pipeline. The run times are dependant on the number of atoms and the grid size. For the molecules we considered in our case studies, we observed running times in the range of 50~ms~(23 atoms) to 150~ms~(97 atoms) on a workstation with an Intel i9 processor having 10 cores. All the grids in our case studies were of similar sizes with around half a million voxels.

\section{Conclusions and future directions}
\label{sec:conclusion} 
In this paper, we have presented a pipeline for visual analysis of electronic transitions by looking at the Natural Transition Orbitals for the hole and particle. 
Our quantitative analysis adds a new perspective to a common problem in quantum chemistry with many applications.
The proposed visual representation provides valuable information when comparing different designs of molecules concerning their physical and chemical properties. 

The proposed pipeline represents a generic approach for applications where segmentation and transitions between those are of interest. However, the individual steps are using application-specific design decisions that have been driven by the goal to provide an efficient, simple, and chemically plausible solution to the problem.   
This is at first the choice of using a weighted Voronoi diagram for the segmentation. Second, the set of constraints in the optimization model are based on chemically plausible conditions. When applying the proposed pipeline to other settings, e.g., when moving from the investigation of molecules to crystal structures, the decisions may need to be reconsidered and adapted to the new application.

We will investigate alternatives for specific design decisions in the future. This includes different geometric, and topological approaches for volume segmentation and a larger variety of geometric constraints in the optimization.  We also plan to extend this work to facilitate the automated classification of electronic transitions of ensembles of molecules. Further, integration of other physical and chemical properties of the transitions like transition energy and oscillatory strength would also be valuable to the chemists.

\appendix

\section{Comparison of Voronoi segmentation with gradient-based segmentation}

For segmenting charge density field spatially among the atoms of the molecule, gradient-based partitioning was suggested by Bader~\cite{Bader1990}. In practice, this idea has been implemented in software like TopoMS~\cite{Bhatia2018} and an implementation by Henkelman \emph{et al.}~\cite{Henkelman2006}. However, both these software failed to generate a segmentation for our input charge density fields \textit{i.e.} NTOs. We believe this is because of the fact that these software are tailored for analysis of \emph{full} charge density fields. The NTO charge density however is different, for example, it may not have any charge density maxima corresponding to some atoms, which is a crucial assumption made by both these software, resulting in failure to generate a segmentation.

In order to compare the Voronoi segmentation with some gradient-based approach, we then decided to use the Morse-Smale complex as implemented in Topology Toolkit~(TTK)~\cite{tierny2017topology}. It uses discrete Morse theory for computation of combinatorial gradient-based segmentation. To compute the segmentation, as a first step the maxima along with their ascending manifolds are computed using TTK. The ascending manifold of a maximum is the set of all points in the domain which reach this maximum after repeated integration in the gradient direction. Then in the second step, we used an approach suggested by TopoMS to assign a maximum to the closest atom, to compute the segmentation and charge per atom. The atomic charge can then be added to compute the subgroup charges.

Table~\ref{table:comparison} lists this detailed comparison for all the data sets used in our case studies. We can observe that Voronoi and gradient-based techniques provide very similar division of charge at the level of subgroups. The instances where the charge computed by Voronoi and Morse complex based approaches differ by more than $2\%$ are highlighted in red. We observed only 6 such cases out of 102 computations of charges at the level of subgroup. Five of these six cases are observed in the case of metal complexes where one of the subgroups consist of just atom which can result in more noticeable differences in atomic boundaries.

\begin{table*}[t]
\begin{center}
\small
\resizebox{1.85\columnwidth}{!}{
%\begin{sideways}
\begin{tabular}{cccccccccc}
    \toprule
    & \multirow{2}{*}{Molecule} & \multirow{2}{*}{State} & \multirow{2}{*}{Subgroup} & \multicolumn{3}{c}{$Q^h$} & \multicolumn{3}{c}{$Q^p$} \\
    \cmidrule(lr){5-7} \cmidrule(lr){8-10}
    & & & & $Q^h_{Vor}$ & $Q^h_{MC}$ & $|Q^h_{Vor}-Q^h_{MC}|$ & $Q^p_{Vor}$ & $Q^p_{MC}$ & $|Q^p_{Vor}-Q^p_{M C}|$ \\
    \midrule
    \multirow{6}{*}{Case study 1} & \multirow{6}{*}{Thiophene-Quinoxaline} & \multirow{2}{*}{State 1} & THIO & 54.8\% & 54.8\% &	\textbf{0.0\%}	& 6.8\% & 5.8\%	& \textbf{1.0\%} \\
    & & & QUIN & 45.2\% & 45.2\% & \textbf{0.0\%} &	93.2\% & 94.2\% & \textbf{1.0\%} \\
    \cmidrule(lr){3-10}
    & & \multirow{2}{*}{State 4} & THIO & 94.2\% & 94.4\% & \textbf{0.2\%} & 7.1\% & 6.0\% & \textbf{1.1\%} \\
    & & & QUIN & 5.8\% & 5.6\% & \textbf{0.2\%} & 92.9\% & 94.0\% & \textbf{1.1\%} \\
    \cmidrule(lr){3-10}
    & & \multirow{2}{*}{State 9} & THIO & 16.5\% & 16.4\% & \textbf{0.1\%} & 2.1\% & 1.8\% & \textbf{0.3\%} \\
    & & & QUIN & 83.5\%	& 83.6\% & \textbf{0.1\%} & 97.9\% & 98.2\% & \textbf{0.3\%} \\
    \midrule
    \multirow{18}{*}{Case study 2} & \multirow{18}{*}{[6]cycloparaphenylene} & \multirow{6}{*}{State 1} & PHE1 & 13.2\% & 13.2\% & \textbf{0.0\%} & 20.3\% & 20.7\% & \textbf{0.4\%} \\
    & & & PHE2 & 15.6\% & 15.6\% & \textbf{0.0\%} & 17.6\% & 17.6\% & \textbf{0.0\%} \\
    & & & PHE3 & 19.2\% & 19.2\% & \textbf{0.0\%} & 14.1\% & 13.8\% & \textbf{0.3\%} \\
    & & & PHE4 & 14.0\% & 14.0\% & \textbf{0.0\%} & 19.2\% & 19.3\% & \textbf{0.1\%} \\
    & & & PHE5 & 20.4\% & 20.4\% & \textbf{0.0\%} & 13.2\% & 13.0\% & \textbf{0.2\%} \\
    & & & PHE6 & 17.6\% & 17.6\% & \textbf{0.0\%} & 15.6\% & 15.6\% & \textbf{0.0\%} \\
    \cmidrule(lr){3-10}
    & & \multirow{6}{*}{State 2} & PHE1 & 0.7\% & 0.5\% & \textbf{0.2\%} & 37.9\% & 39.7\% & \textbf{1.8\%} \\
    & & & PHE2 & 0.7\% & 0.5\% & \textbf{0.2\%} & 38.0\% & 39.4\% & \textbf{1.4\%} \\
    & & & PHE3 & 11.4\% & 11.4\% & \textbf{0.0\%} & 10.5\% & 8.9\% & \textbf{1.6\%} \\
    & & & PHE4 & 11.4\% & 11.4\% & \textbf{0.0\%} & 10.5\% & 9.2\% & \textbf{1.3\%}\\
    & & & PHE5 & 37.9\% & 38.0\% & \textbf{0.1\%} & 1.5\% & 1.4\% & \textbf{0.1\%} \\
    & & & PHE6 & 37.9\%	& 38.0\% & \textbf{0.1\%} & 1.5\% & 1.4\% & \textbf{0.1\%} \\
    \cmidrule(lr){3-10}
    & & \multirow{6}{*}{State 3} & PHE1 & 3.3\% & 3.2\%	& \textbf{0.1\%} & 24.1\% & 23.6\% & \textbf{0.5\%} \\
    & & & PHE2 & 24.8\% & 24.9\% & \textbf{0.1\%} & 3.1\% & 2.6\% & \textbf{0.5\%} \\
    & & & PHE3 & 43.3\% & 43.4\% & \textbf{0.1\%} & 1.8\% & 1.8\% & \textbf{0.0\%} \\
    & & & PHE4 & 0.5\% & 0.4\% & \textbf{0.1\%} & 43.8\% & 46.1\% & \textbf{\textcolor{red}{2.3\%}} \\
    & & & PHE5 & 24.8\% & 24.9\% & \textbf{0.1\%} & 3.1\% & 2.6\% & \textbf{0.5\%} \\
    & & & PHE6 & 3.3\% & 3.2\% & \textbf{0.1\%} & 24.1\% & 23.3\% & \textbf{0.8\%} \\
    \midrule
    \multirow{9}{*}{Case study 3} & \multirow{3}{*}{Cu-PHE2} & \multirow{3}{*}{State 1} & Cu & 71.6\% & 70.3\% & \textbf{1.3\%} & 3.5\% & 1.8\% & \textbf{1.7\%} \\
    & & & PHE1 & 13.8\% & 14.3\% & \textbf{0.5\%} & 42.7\% & 43.4\% & \textbf{0.7\%} \\
    & & & PHE2 & 14.6\% & 15.4\% & \textbf{0.8\%} & 53.8\% & 54.8\% & \textbf{1.0\%} \\
    \cmidrule(lr){2-10}
    & \multirow{3}{*}{Ag-PHE2} & \multirow{3}{*}{State 1} & Ag & 52.3\% & 49.8\% & \textbf{\textcolor{red}{2.5\%}} & 2.7\% & 0.7\% & \textbf{2.0\%} \\
    & & & PHE1 & 23.0\% & 24.3\% & \textbf{1.3\%} & 46.8\% & 47.8\% & \textbf{1.0\%} \\
    & & & PHE2 & 24.7\% & 25.9\% & \textbf{1.2\%} & 50.5\% & 51.5\% & \textbf{1.0\%} \\
    \cmidrule(lr){2-10}
    & \multirow{3}{*}{Au-PHE2} & \multirow{3}{*}{State 1} & Au & 50.2\% & 49.1\% & \textbf{1.1\%} & 2.8\% & 1.3\% & \textbf{1.5\%} \\
    & & & PHE1 & 29.8\% & 30.4\% & \textbf{0.6\%} & 46.9\% & 47.6\% & \textbf{0.7\%} \\
    & & & PHE2 & 20.0\% & 20.6\% & \textbf{0.6\%} & 50.3\% & 51.1\% & \textbf{0.8\%} \\
    \midrule
    \multirow{18}{*}{Case study 4} & \multirow{3}{*}{Cu-PHE2} & \multirow{3}{*}{State 1} & Cu & 71.6\% & 70.3\% & \textbf{1.3\%} & 3.5\% & 1.8\% & \textbf{1.7\%} \\
    & & & PHE1 & 13.8\% & 14.3\% & \textbf{0.5\%} & 42.7\% & 43.4\% & \textbf{0.7\%} \\
    & & & PHE2 & 14.6\% & 15.4\% & \textbf{0.8\%} & 53.8\% & 54.8\% & \textbf{1.0\%} \\
    \cmidrule(lr){2-10}
    & \multirow{3}{*}{Cu-PHE-PHEPHE} & \multirow{3}{*}{State 1} & Cu & 67.9\% & 66.5\% & \textbf{1.4\%} & 3.2\% & 1.7\% & \textbf{1.5\%} \\
    & & & PHE & 12.9\% & 13.7\% & \textbf{0.8\%} & 59.2\% & 60.5\% & \textbf{1.3\%} \\
    & & & PHEPHE & 19.2\% & 19.8\% & \textbf{0.6\%} & 37.6\% & 37.8\% & \textbf{0.2\%} \\
    \cmidrule(lr){2-10}
    & \multirow{3}{*}{Cu-PHE-PHEME} & \multirow{3}{*}{State 1} & Cu & 70.9\% & 69.9\% & \textbf{1.0\%} & 3.6\% & 2.0\% & \textbf{1.6\%} \\
    & & & PHE & 16.9\% & 18.0\% & \textbf{1.1\%} & 95.7\% & 97.3\% & \textbf{1.6\%} \\
    & & & PHEME & 12.2\% & 12.1\% & \textbf{0.1\%} & 0.7\% & 0.6\% & \textbf{0.1\%} \\
    \cmidrule(lr){2-10}
    & \multirow{3}{*}{Cu-PHE-PHEOME} & \multirow{3}{*}{State 1} & Cu & 66.5\% & 65.8\% & \textbf{0.7\%} & 3.2\% & 1.9\% & \textbf{1.3\%} \\
    & & & PHE & 14.7\% & 15.5\% & \textbf{0.8\%} & 95.8\% & 97.4\% & \textbf{1.6\%} \\
    & & & PHEOME & 18.8\% & 18.7\% & \textbf{0.1\%} & 1.1\% & 0.7\% & \textbf{0.4\%} \\
    \cmidrule(lr){2-10}
    & \multirow{3}{*}{Cu-PHE-IPR} & \multirow{3}{*}{State 1} & Cu & 71.4\% & 70.8\% & \textbf{0.6\%} & 4.1\% & 1.5\% & \textbf{\textcolor{red}{2.6\%}} \\
    & & & PHE & 23.0\% & 24.9\% & \textbf{1.9\%} & 93.8\% & 97.2\% & \textbf{\textcolor{red}{3.4\%}} \\
    & & & IPR & 5.6\% & 4.3\% & \textbf{1.3\%} & 2.0\% & 1.3\% & \textbf{0.7\%} \\
    \cmidrule(lr){2-10}
    & \multirow{3}{*}{Cu-PHE-XANT} & \multirow{3}{*}{State 1} & Cu & 35.1\% & 30.6\% & \textbf{\textcolor{red}{4.5\%}} & 2.3\% & 1.2\% & \textbf{1.1\%} \\
    & & & PHE & 8.0\% & 7.9\% & \textbf{0.1\%} & 95.5\% & 97.1\% & \textbf{1.6\%} \\
    & & & XANT & 56.9\% & 61.5\% & \textbf{\textcolor{red}{4.6\%}} & 2.2\% & 1.7\% & \textbf{0.5\%} \\
    \bottomrule
\end{tabular}
}
\end{center}
\caption{Comparison of subgroup charges computed using the Voronoi-based approach and the Morse complex-based approach. The symbols $Q^h_{Vor}$ and $Q^h_{MC}$ are used to denote the subgroup charge for hole NTO computed using Voronoi diagram-based and Morse complex-based approaches. Similarly, $Q^p_{Vor}$ and $Q^p_{MC}$ are used for particle NTO charges. The entries where the two approaches differ by more than $2\%$ are highlighted in red.}
\label{table:comparison}
\end{table*}

\small{
\noindent{\textbf{Acknowledgements.}} This work is supported by the SeRC (Swedish e-Science Research Center), the Swedish Research Council~(VR) grant 2019-05487, and an Indo-Swedish joint network project: DST/INT/SWD/VR/P-02/2019 and VR grant 2018-07085. VN is partially supported by a Swarnajayanti Fellowship from the Department of Science and Technology, India (DST/SJF/ETA-02/2015-16) and a Mindtree Chair research grant. The computations were enabled by resources provided by the Swedish National Infrastructure for Computing (SNIC) at NSC partially funded by the VR grant agreement no. 2018-05973.
}

\clearpage
%-------------------------------------------------------------------------
% bibtex
\bibliographystyle{eg-alpha-doi}  
\bibliography{references}        

% biblatex with biber
%\printbibliography

\end{document}